\documentclass[reprint, jcp,  superscriptaddress]{revtex4-2}

\DeclareMathAlphabet{\mathcal}{OMS}{cmsy}{m}{n}

\newcommand{\be}{\begin{equation}}
\newcommand{\ee}{\end{equation}}

\newcommand{\mean}[1]{\langle #1 \rangle}

\newcommand{\bea}{\begin{eqnarray}}
\newcommand{\eea}{\end{eqnarray}}
 
\newcommand{\ba}{\begin{array}}
\newcommand{\ea}{\end{array}}

\newcommand{\bl}{\begin{flalign}}
\newcommand{\enl}{\end{flalign}}

\usepackage{verbatim}
\usepackage{graphicx}
\usepackage{color}
\usepackage{amsmath}
\usepackage{amssymb}
\usepackage{braket}
\usepackage{enumerate}
\usepackage{multirow}
\usepackage{colortbl}
\definecolor{Gray}{gray}{0.3}
\usepackage{wrapfig}
\usepackage{makecell}

\begin{document}

\title{Microscopic Theory, Analysis,  and Interpretation of Conductance Histograms in Molecular Junctions}

\author{Leopoldo Mej\'ia}
\email{leopoldo.mejia@berkeley.edu}
\affiliation{
    Department of Chemistry, University of Rochester, Rochester, New York 14627, USA
    }
\altaffiliation{Current Address: Department of Chemistry, University of California, Berkeley, California 94720, USA} 
\altaffiliation{Materials Sciences Division, Lawrence Berkeley National Laboratory, Berkeley, California 94720, USA} 
\author{Pilar Cossio} 
\affiliation{Center for Computational Mathematics, Flatiron Institute, New York City, NY 10010, USA}
\affiliation{Center for Computational Biology, Flatiron Institute, New York City, NY 10010, USA}
\author{Ignacio Franco}
\email{ignacio.franco@rochester.edu}
\affiliation{
    Department of Chemistry, University of Rochester, Rochester, New York 14627, USA
    }
\affiliation{
    Department of Physics, University of Rochester, Rochester, New York 14627, USA
    }


\begin{abstract}
 Molecular electronics break-junction experiments  are widely used to investigate fundamental physics and chemistry at the nanoscale. Reproducibility in these experiments relies on measuring conductance on thousands of freshly formed molecular junctions, yielding a broad histogram of conductance events. Experiments typically focus on the most probable conductance, while the information content of the conductance histogram has remained unclear. Here, we develop a theory for the conductance histogram,  which accurately fits experimental data and augments the information content that can be extracted, by merging the theory of force-spectroscopy with molecular conductance. Specifically, we propose a microscopic model of the junction evolution under the modulation of external mechanical forces and combine it with the non-equilibrium stochastic features of junction rupture and formation. Our formulation captures contributions to the conductance dispersion that emerge due to changes in the conductance during the mechanical elongation inherent to the experiments. The final histogram shape is determined by the statistics of junction rupture and formation. The procedure yields analytical equations for the conductance histogram in terms of parameters that describe the free-energy profile of the junction, its mechanical manipulation, and the ability of the molecule to transport charge.  All physical parameters that define our microscopic model can be obtained from separate conductance and rupture force measurements on molecular junctions. Further, the predicted behavior with respect to physical parameters can be used to test the range of validity of the microscopic model, understand the conductance histograms, design molecular junction experiments with enhanced resolution and molecular devices with more reproducible conductance properties.
\end{abstract}

\maketitle

The study of charge transport across single-molecules is a powerful tool to investigate fundamental physics and chemistry at the nanoscale\cite{elke2017molecular, datta2005quantum, nitzan2006chemical, coropceanu2007charge, nitzan2003electron, bergfield2013forty,cabosart2019reference}. In particular, single-molecule conductance measurements have been used to investigate conformational dynamics\cite{mejia2018signatures, venkataraman2006dependence, mishchenko2010influence, wu2020folding}, chemical reactions\cite{mejia2019force, mejia2021diels, li2017mechanical, aragones2016electrostatic, guan2018direct, huang2017single}, quantum interference\cite{ballmann2012experimental, guedon2012observation, garner2018comprehensive}, charge transport coherence\cite{mejia2022coherent} and to develop single-molecule spectroscopies\cite{pirrotta2017single, koch2018structural}.
Further, they are routinely used to establish structure-transport relations that can guide the design and our ability to understand photovoltaics\cite{nelson2009modeling, germack2009substrate, breeze2001charge}, redox catalysis\cite{yagi2006charge}, energy transport and storage\cite{heeger201425th}, photosynthesis\cite{golbeck1992structure}, and biological signaling\cite{sontz2012dna}. In addition, the platform has been used to construct molecular-based devices such as switches\cite{Franco-2011,wu2020folding}, transistors\cite{li2016proton, ghosh2004gating, lang2005charge, fathizadeh2018engineering}, and diodes\cite{diez2009rectification, elbing2005single}.

\begin{figure*}[t]
 \centering
\includegraphics[scale=1.0]{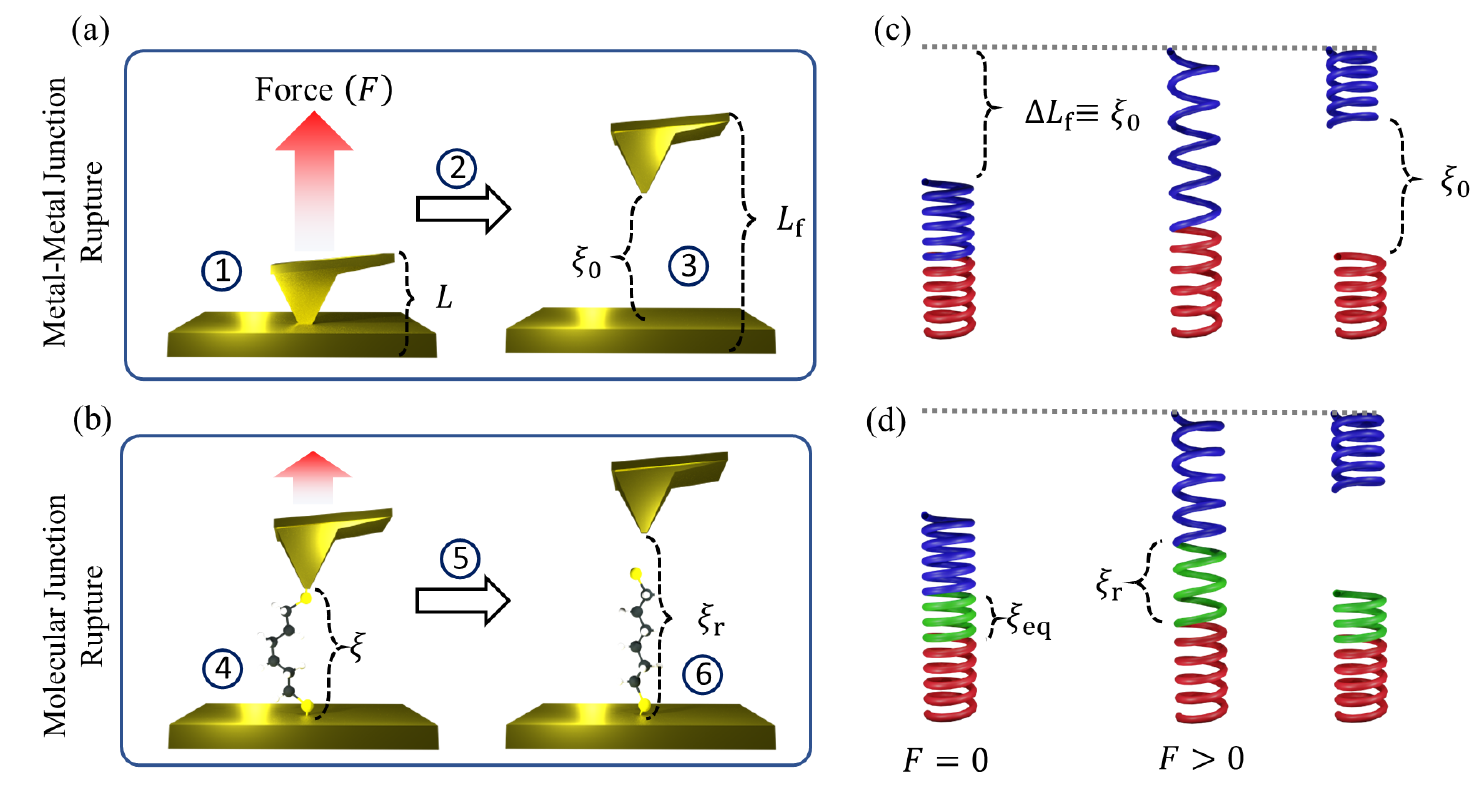}
\caption{Break-junction experiments. (a) The rupture of the metal-metal junction leads to an initial electrode-electrode gap ($\xi_0$) in which the molecule is anchored forming a molecular junction. The circled numbers signal the steps into which the process has been divided, as described in Sec.~\ref{sec:theory}. (b) The pulling of the molecular junction results in its rupture at electrode gap $\xi_\text{r}$. Both $\xi_0$ and $\xi_\text{r}$ are stochastic variables determined by rupture statistics. (c) The metal-metal rupture can be seen as the rupture of two brittle springs connected in series. The blue and red springs represent the electrodes (surface and cantilever in a Scanning Tunneling Microscope break-junction experiment). (d) The molecular junction rupture can be represented by the rupture of three brittle springs connected in series, where the green spring represents the molecule.}
\label{springs}
\end{figure*}

A common scheme to measure single-molecule conductance is the so-called break-junction setup\cite{xu2003measurement, he2006measuring, widawsky2012simultaneous, konishi2013single, venkataraman2006single, zhou2008extending}. In these experiments (see Fig. \ref{springs}) two metallic electrodes are brought into mechanical contact and then pulled apart until a nanoscale gap forms between them due to the rupture of the metal-metal junction. Molecules in the surrounding medium bridge the gap between the two electrodes by attaching their ends to the metallic contacts, which results in the formation of a molecular junction. As the formed molecular junction is elongated by mechanically pulling, a voltage is applied and the resulting current is recorded. The pulling is continued until the molecular junction ruptures. This process is repeated thousands of times on freshly formed junctions and the distribution of conductance events, the conductance histogram, is reported. 

While the conductance of individual molecular junctions is challenging to experimentally reproduce, the conductance histogram is highly reproducible. Nevertheless, these histograms typically exhibit a broad conductance dispersion of $\sim$ 0.5-2 orders of magnitude with respect to the most probable conductance value\cite{li2020understanding, gonzalez2006electrical, li2019molecular, quek2009mechanically, zhou2008extending, cabosart2019reference}. This limits the utility of break-junction techniques as a platform to investigate single-molecules and construct molecular-based devices, as the broad conductance features imposes fundamental limits on the resolution of individual molecular events and the design of devices with reproducible conductance properties. 

To extract physical information from the conductance histograms, and design useful strategies to narrow their width, it is desirable to develop a microscopic theory of the conductance distribution in break-junction experiments. Such a theory could be used to interpret and predict the role that external factors --such as the pulling speed, cantilever stiffness, and temperature--, and internal molecular features --such as the molecular structure and chemical anchor groups--, play in determining the width and shape of the conductance histograms. Further, the theory could help bridge the gap between atomistic simulations of molecular conductance that typically focus on few representative junction conformations and measurements that record all statistically possible experimentally accessible events, and thus require a statistical approach\cite{li2019molecular,tschudi2016estimating}.

However, the complexity of the microscopic origin of the conductance dispersion in molecular junctions has prevented the development of such a theory. Specifically, it has been shown that multiple factors, including changes in the molecular conformation\cite{mejia2018signatures, Paulsson2009a}, various electrode-molecule binding configurations\cite{Li2006, Li2008a}, variations in the electrode geometry\cite{French2013b,French2012}, and the systematic mechanical manipulation of the junction\cite{li2020understanding} broaden the conductance histogram, making it challenging to formulate a theory in a unified framework.

To make systematic progress, it is necessary to focus on the contributions of a few microscopic factors to the conductance dispersion. In this regard, a recent theory-experiment analysis\cite{li2020understanding} revealed that a broad conductance histogram will still emerge even in ideal experiments where the electrode geometry and molecular binding configuration can be perfectly controlled. Specifically, it was observed that the changes in conductance due to the mechanical manipulation of the junction alone already account for the observed conductance dispersion in break-junction experiments.  In addition, the study in Ref.\cite{li2020understanding} showed that one of the reasons why these experiments require collecting statistics is because forming and breaking the junction is an inherently stochastic process that needs to be sampled to generate reproducible histograms.

Here, we propose a microscopic theory of conductance histograms by considering the mechanical manipulation of the junction and the statistics of junction formation and rupture as the only sources of conductance dispersion. The theory describes the conductance histograms in terms of physically meaningful parameters that characterize the free-energy profile (FEP) of the junction, its mechanical manipulation, and the ability of the molecule to transport charge. While this view of the origin of the conductance histogram is only capturing one of the possible contributing factors to the conductance dispersion, it allows us to obtain analytic expressions that are useful for fitting, modeling, and interpreting experimental conductance histograms.

This study complements and advances previous efforts to model conductance histograms through phenomenological broadening of junction parameters\cite{quan2015quantitative, reuter2012signatures, williams2013level} and molecular dynamics simulations of junction formation and evolution\cite{li2019molecular, li2020understanding, deffner2022learning}, and efforts to classify molecular conductance events through machine learning\cite{cabosart2019reference, el2019unravelling, van2022benchmark}. The main advance is that it provides a useful microscopic picture of junction formation and evolution that recovers the shape of experimental conductance histogram, enables extracting information about the microscopic parameters and establishes a foundation for generalizations.

Specifically, inspired by force-spectroscopy experiments for protein unfolding and unbinding processes\cite{cossio2016kinetic, hummer2003kinetics, cossio2015artifacts, dudko2006intrinsic, hyeon2012multiple, evans1997dynamic, bell1978models}, we develop a theory of break-junction experiments in which the metal-metal and metal-molecule rupture events, that lead to the formation and breaking of the molecular junction, are modeled as stochastic escapes from one-dimensional wells modulated by mechanical forces. This leads to a range of initial and final gaps between electrodes (electrode gap) that determine the  molecular ensemble that is sampled in the experiment. By relating the electrode gap to junction conductance we isolate an expression for the conductance histogram. The main results of this work are analytical equations (Eqs. \ref{PlogT_gen} and \ref{PlogT}) that can properly capture the shape and peak position of experimental conductance histograms --including aliphatic, aromatic and radical containing molecules with varying anchor groups, and supramolecular complexes-- and that are defined by microscopic parameters that capture the electro-mechanical properties of the junction. 
The theory can be used to understand how the histograms change with molecular design and experimental conditions such as pulling speed, junction stiffness, and temperature.  Importantly, all microscopic features that define the conductance histogram can be extracted by supplementing the conductance break-junction experiments with force spectroscopy of junction rupture, thus providing a general platform to augment the information content that can be extracted from this class of experiments.


\section{\label{sec:theory}Theory}

\begin{table*}[t]
\centering{}\caption{Physical parameters defining the theory of conductance histograms. Example values (column 3) were used to construct Figs. \ref{hist_evol} and \ref{hist_param}. Parameters  in column 4 were recovered  from fitting synthetic rupture force and conductance histograms, generated with the example parameters in column 3, to Eqs. \ref{pF} and \ref{PlogT} (standard deviation in brackets).  The "f" and "r" symbols in brackets refer to the rupture of the metal-metal and molecular junction, respectively. The quantities $\beta$, $\dot{F}$, $\kappa$ and $\kappa_\xi$  are experimentaly known. All other parameters  can be extracted by fitting experimental data to the theory.}\label{tab0}
\begin{tabular}{cccc}
\hline
Symbol &  Meaning &  Example & \\
\hline
$\beta$      & Inverse temperature  & 38.68 eV$^{-1}$ (300K) & \\ 

$\dot{F}$    & Loading rate         & \makecell{400 nN s$^{-1}$ (f) \\ 400 nN s$^{-1}$ (r)} & \\
$\kappa$     & Junction spring constant      & \makecell{8 N m$^-1$ (f) \\ 8 N m$^-1$ (r)} & \\

$\kappa_{\xi}$ & Molecular spring constant & 8 N m$^-1$ & \\

\hline

& & & Fitted (SD) \\

$\xi_\text{eq}$   & \makecell{Molecular junction electrode gap at mechanical equilibrium} & 1.50 \AA & 1.35 (0.05) \AA\\

$\chi^\ddag$ & \makecell{Distance between \\$\xi_\text{eq}$ and transition state in the FEP} & \makecell{0.200 \AA{} (f) \\ 0.200 \AA{} (r)} & \makecell{0.198 (0.002) \AA{} (f) \\ 0.199 (0.002) \AA{} (r)}\\

$k_0$        & Spontaneous rupture rate ($F=0$) & \makecell{1.00 s$^{-1}$ (f) \\ 20.00 s$^{-1}$ (r)} & \makecell{1.30 (0.08) s$^{-1}$ (f) \\ 24.81 (1.12) s$^{-1}$ (r)}\\

$\gamma$     & Transmission decay coefficient & -1.15 \AA$^{-1}$ & -1.14 (0.01) \AA$^{-1}$\\
$\log T_0$        & Base transmission ($\xi\to 0$) & -3.00 & -2.97 (0.02)\\

\hline
\end{tabular}
\end{table*}

To develop a theory for the conductance histograms, we partition the break-junction experiment (Fig. \ref{springs}) into six main events: (1) the formation of a contact between the two metallic electrodes (Fig. \ref{springs}a, left); (2) the mechanical elongation of the metal-metal contact and (3) its rupture to create a nanoscopic gap (Fig. \ref{springs}a, right); (4) the attachment of a molecule bridging this gap between the two electrodes that is (5) subsequently mechanically elongated (Fig. \ref{springs}b, left) until (6) junction rupture (Fig. \ref{springs}b, right). Processes (3) and (6) are stochastic, thermally activated, and nonequilibrium in nature. Our view is that, because of this,  to recover reproducible conductance features it is necessary to statistically sample all possible rupture events by repeating the experiment (steps (1)-(6)) thousands of times.

In the theory, the conductances that enter into the histograms are those encountered by the junction during (5). The distribution of junction elongations that determine such conductances are given by the distribution of initial nanoscopic gaps in (3) and at rupture in (6). The probability of a given conductance value, thus, depends on the probability that a given electrode gap  is visited during (5) and the  relation between molecular junction conductance and such electrode gaps.

The theory supposes that there is an effective one-to-one relation between measured conductance $\langle G(\xi)\rangle $ and junction gap $\xi$. For a given gap, there is a whole thermal ensemble of possible molecular and junction configurations that are accessible and contribute to the conductance. However, since experiments measure a  current that is \emph{time-averaged} over microseconds, these individual contributions are averaged out and cannot be experimentally resolved leading to simpler conductance traces. One coordinate that systematically changes during pulling in timescales slower than the integration time of the current is the electrode gap, $\xi$. In experiments, the junction is pulled with speeds of  nm/s, and thus sub-\AA ngstrom variations of electrode gap $\xi$ can be experimentally resolved in the conductance measurements.\cite{li2020understanding} This effectively leads to a conductance that parametrically depends on the electrode gap, $\langle G(\xi)\rangle$.

We note that, even in the presence of time-averaging, the experiments can discriminate molecular conformations that are mechanically stabilized, or that survive for times longer than the current integration time, such as changes in the binding configuration or transitions between stable  molecular conformations (e.g. gauche vs. trans isomers in alkanes or mechanically activated reactions). The one-to-one assumption for $\langle G(\xi)\rangle$ accounts for many of these changes, but it cannot capture physical situations in which multimodal conductance features are accessible at a given electrode gap. When there is no interconversion between these modes, multimodality can be simply accounted for through independent one-to-one $\langle G(\xi)\rangle$ relations. The proposed theory can also be generalized to the more complex case where different long-lived conformers are accessible at a given $\xi$ and interconvert during the pulling process. For definiteness, we focus on a one-to-one relation for $\langle G(\xi)\rangle$ and discuss its generalization in the outlook.

The experimentally controllable variable is  the overall length of the junction $L$ and not $\xi$ (Fig.~\ref{springs}). These quantities do not coincide as during pulling there can be mechanical deformations of the electrodes that do not lead to changes in the electrode gap. In experiments, it is customary to align different conductance traces $\mean{G(L)}$ at their rupture point. 
In Ref. \cite{li2020understanding} we showed computationally that even in ideal experiments (with reproducible electrode shape and binding configuration) this will lead to a dispersion of $\mean{G(L)}$ curves as the elongation in which the junction ruptures varies between experiments.  However, as shown in  Ref. \cite{li2020understanding}, if the conductance trajectories are plotted with respect to electrode gap they collapse into a single curve, justifying the one-to-one relation $\mean{G(\xi)}$ in the theory.

To describe the nonequilibrium stochastic features inherent to the junction formation and rupture, we take advantage of the theory of rupture-force statistics developed in biophysics to describe mechanically modulated transitions between two states \cite{cossio2016kinetic, hummer2003kinetics, cossio2015artifacts, dudko2006intrinsic, hyeon2012multiple, evans1997dynamic, bell1978models} as needed, for instance, in the description of the force-spectroscopy of protein unfolding. To map this into conductance, which is the main observable, we first develop a model for the mechanical driving of the molecular junction that maps the distribution of rupture forces into a distribution  of electrode gaps at junction formation and rupture. Then, based on the parametric relation between the average conductance and electrode gap $\mean{G(\xi)}$, we connect the distribution of initial and final junction elongations to a distribution of conductances. This results in a general equation for the conductance histogram. We further specialize the model to the case in which the conductance depends exponentially on $\xi$ that is then used to fit representative experiments, and analyze the theory. For clarity in presentation, Table \ref{tab0} summarizes the symbols and physical meaning of the parameters of the theory.


\subsection{Probability density function of rupture-forces}

\begin{figure}[t]
 \centering
\includegraphics[scale=1.0]{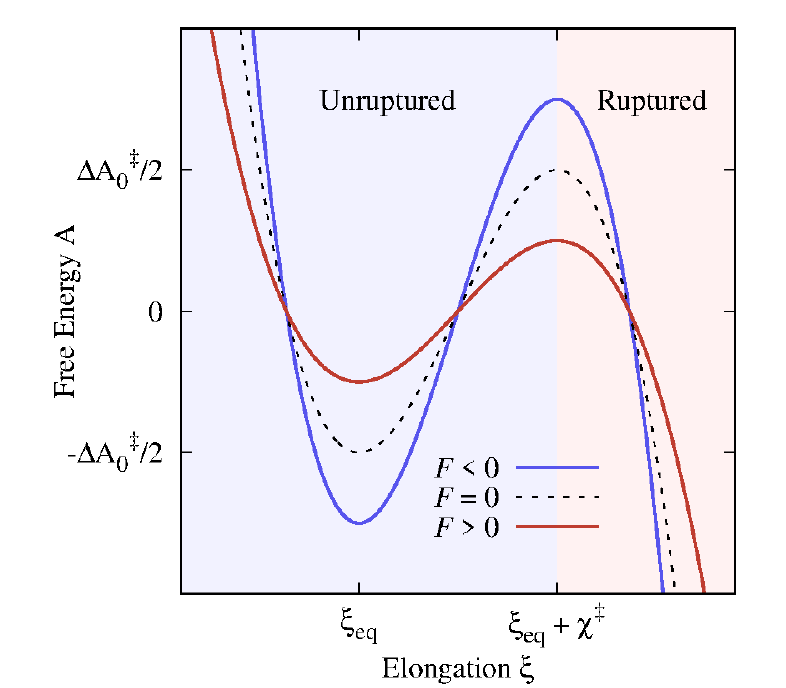}
\caption{Schematic representation of the free-energy profile (FEP) of a metal-metal or molecular junction along the pulling coordinate. External mechanical forces ($F$) decrease (red line, $F>0$) or increase (blue line, $F<0$) the free-energy barrier between the unruptured and ruptured states ($\Delta A^{\ddag}$).}
\label{bell_PES}
\end{figure}

The rupture of the metal-metal contact and of the molecular junction (processes (3) and (6)) can be understood as a free energy barrier-crossing event in the presence of an external force from  an unruptured state to a ruptured state. Specifically, the thermodynamic state of the (metal-metal or molecular) junction is represented by a one-dimensional FEP along the pulling coordinate (see Fig. \ref{bell_PES}). The unruptured state corresponds to the bottom-well in the FEP with equilibrium elongation $\xi_\text{eq}$ at zero force ($F=0$). The transition state between the ruptured and unruptured state, represented by the barrier top, is located a distance $\chi^\ddag$ away from the well bottom, and has a $\Delta A^\ddag_0$ activation free energy at $F=0$. The unruptured state corresponds to the unbounded portion of the FEP ($\xi>\xi_{\text{eq}}+\chi^\ddag$).

The application of an external force modifies the FEP, modulating the junction rupture rates. For example, positive forces ($F>0$) reduce the activation barrier making it more probable for the junction to rupture. We assume a brittle system where $\xi_\text{eq}$ and the distance to the transition state ($\chi^{\ddag}$) do not change with force as shown in Fig. \ref{bell_PES}. In this mechanically brittle limit\cite{cossio2016kinetic}, the force-dependent rupture rate follows Bell's formula\cite{bell1978models}:
\begin{equation}
\label{bell}
    k(t) = k_0\mathrm{e}^{\beta\chi^\ddag F(t)},
\end{equation}
where $k_0$ is the spontaneous rupture rate at $F=0$, $\beta$ is the inverse temperature and where the force $F(t) = \dot{F} t$ is assumed to increase linearly in time with a constant loading rate $\dot{F}$.  Equation \ref{bell} implies that the activation energy $\Delta A^{\ddag} = \Delta A_0^\ddag - F(t)\chi^\ddag$ varies linearly with $F(t)$, as represented in Fig. \ref{bell_PES} for positive (pulling) and negative (pushing) forces. This was shown by Bell~\cite{bell1978models} for soft pulling springs by considering that the FEP is distorted in the presence of forces as $A(\chi, t) = A_0(\chi)+  \kappa(\chi -(\dot{F}/\kappa) t)^2/2$ and using Arrhenius formula.
Under the assumption that the survival probability of the junction $S(t)$ follows a first-order rate equation of the form $\dot{S}(t)=-k(t)S(t)$ and using Bell's formula (Eq. \ref{bell}) for the rate coefficient, the probability density function of rupture forces $p_F(F)$ --i.e., the force required to mechanically break the junction-- is (see Ref. \citenum{hummer2003kinetics} for details):
\begin{equation}
\label{pF}
    p_F(F)=\frac{k_0}{\dot{F}}\exp{\Big[ \beta\chi^\ddag F - \frac{k_0}{\dot{F}\beta\chi^\ddag} \mathrm{e}^{\beta\chi^\ddag F} \Big]}.
\end{equation}

Equation \ref{pF} was first obtained by Schulten \textit{et al.}\cite{izrailev1997molecular} to describe the dynamics of the unbinding of the Avidin-Biotin complex, and has been widely used to investigate the kinetics of single-molecule pulling experiments in the context of biophysics\cite{hummer2003kinetics, chen2001selectin, bell1978models, alon1995lifetime, tees1993interaction, schmidt2012single, lin2007bell}. Its accuracy depends on the range of validity of Bell's approximation. It has been found that Bell's approach is accurate in the low-force regime\cite{vullev2005modulation} where the applied force does not completely deplete the activation free-energy barrier. 

In the context of molecular electronics Scanning Tunneling Microscopy break-junction (STM-BJ) experiments, the rupture force of a metal-metal \cite{pobelov2017dynamic} and molecular \cite{huang2006measurement} junction has well-defined experimentally-accessible regimes where it satisfies Bell's formula. Both junctions also show a regime where the rupture force becomes independent of pulling rate that is beyond the regime of applicability of the theory.
Our analysis pertains to experiments performed under conditions in which Bell's theory is applicable.


\subsection{Molecular junction gaps along pulling}\label{sec.elong}

Break-junction experiments involve two junction rupture events during the mechanical pulling: the rupture of the metal-metal contact to create the gap in which the molecule is initially placed (Fig. \ref{springs}a) and the rupture of the formed molecular junction (Fig. \ref{springs}b). Each of these rupture events has an associated probability density function of rupture forces determined by Eq. \ref{pF}.

As represented in Fig. \ref{springs}c, the pulling of the metal-metal junction is analogous to the pulling of two brittle springs connected in series. The bottom spring (red in Fig. \ref{springs}c) represents the deformation of the bottom electrode surface with elasticity constant $\kappa_\text{surf}$. The top spring (blue in Fig. \ref{springs}c) represents the deformation of a cantilever or top electrode (whichever is softer) with spring constant $\kappa_\text{cant}$. The effective spring constant of the composite system ($\kappa_\text{f}$) is then given by $1/{\kappa_\text{f}} = 1/{\kappa_\text{surf}} +  1/{\kappa_\text{cant}}$.

The junction is elongated $\Delta L>0$ until its rupture at $\Delta L = \Delta L_\text{f}$. At this point the electrodes return to their mechanical equilibrium leaving a gap $\Delta L_\text{f} = \xi_0$, where we have assumed that there is no plastic deformation of the gold electrodes, i.e. changes in the electrode geometry due to the mechanical \cite{kruger2002pulling, wang2015gold, wang2021decoding, deffner2023learning}. Such events change the equilibrium length of the electrodes but leave the analysis intact. Further extensions of the model that allow transitions from brittle to ductile regimes can be captured by adding an additional parameter to the free energy profile, as proposed in Ref. \cite{cossio2016kinetic}.

The distribution of metal-metal rupture forces determines the distribution of initial $\xi_0$ gaps. Specifically, a rupture force $F_\text{f}=\kappa_\text{f}\Delta L_\text{f}=\kappa_\text{f}\xi_0$ leads to a $\xi_0 = \frac{F_\text{f}}{\kappa_\text{f}}$ gap. Therefore, the probability density function of initial electrode-electrode gaps, $p_0(\xi_0)$, can be calculated from the probability density function of rupture forces $p_F(F)$, as $p_0(\xi_0) =\kappa_\text{f} p_F(F=\kappa_\text{f} \xi_0)$.

Then, from Eq. \ref{pF} it follows that
\begin{equation}
\label{p0}
    p_0(\xi_0)=\frac{k_{0\text{f}}\kappa_\text{f}}{\dot{F}_\text{f}}\exp{\Big[ \beta\chi_\text{f}^\ddag \kappa_\text{f}\xi_0 - \frac{k_{0\text{f}}}{\dot{F}_\text{f}\beta\chi_\text{f}^\ddag} \mathrm{e}^{\beta\chi_\text{f}^\ddag \kappa_\text{f}\xi_0} \Big]}~.
\end{equation}
All quantities in Eq. \ref{p0} refer to the metal-metal rupture and its FEP, as signaled by the ``$\text{f}$" (junction formation) subscript throughout.

\begin{figure*}[t]
 \centering
\includegraphics[scale=1.0]{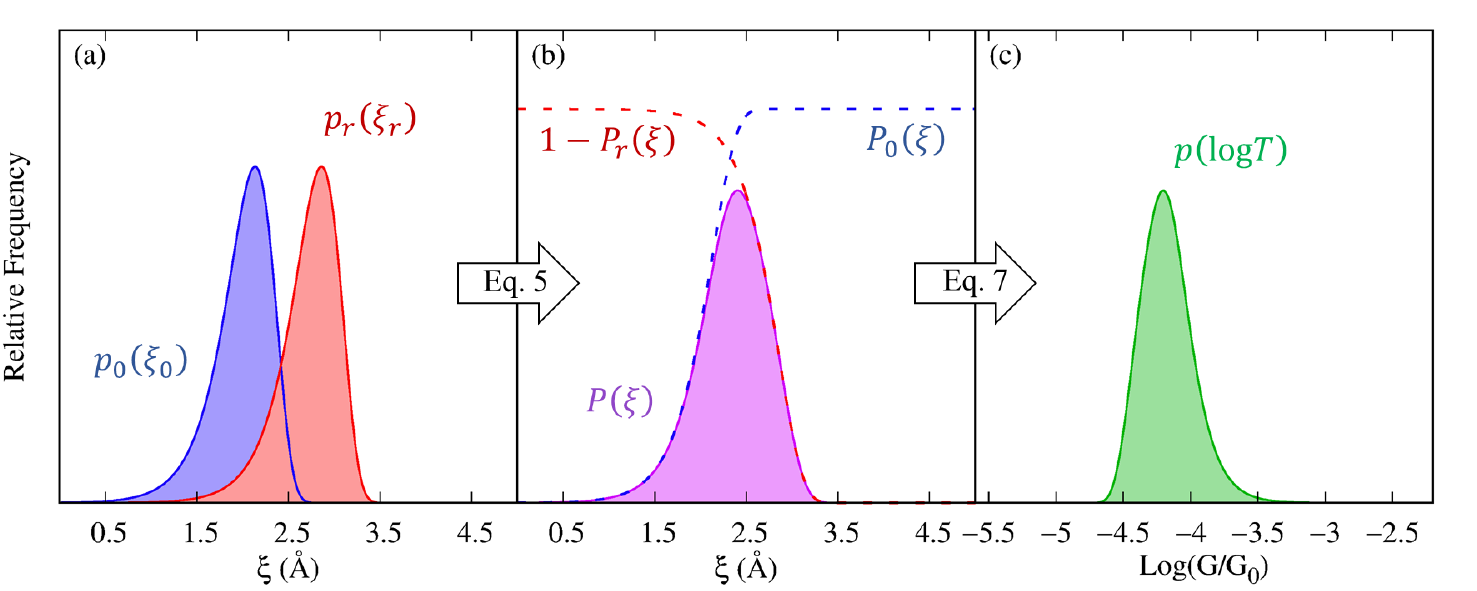}
\caption{Modelling of conductance histogram in a break-junction experiment. (a) Probability density function of the initial ($p_0(\xi_0)$, Eq. \ref{p0}) and rupture ($p_\text{r}(\xi_\text{r})$, Eq. \ref{pr}) electrode gaps  in a molecular junction. (b) Probability of visiting the electrode gap $\xi$ ($P(\xi)$) during a break-junction experiment. The dotted lines represent the probability that the junction has been formed ($P_0(\xi)$)/has not been ruptured ($1-P_\text{r}(\xi)$) at a given $\xi$. (c) Conductance histogram calculated with Eq. \ref{PlogT}. In all cases, the parameters in Table \ref{tab0} were used.}
\label{hist_evol}
\end{figure*}

Similarly, the probability density function of molecular junction gaps at rupture ($\xi_\text{r}$) are determined by the probability density function of rupture forces of the molecular junction. As proposed in Fig. \ref{springs}d, the mechanical response of the molecular junction can be viewed as three brittle springs connected in series. When a force is applied to the combined spring, the same force is applied to each individual spring. Thus, the exerted force at rupture $F_\text{r}=\kappa_\text{r}\Delta L_\text{r}$, where $\kappa_\text{r}$ is the overall spring constant and $\Delta L$ the overall elongation, can be written in terms of the gap between electrodes $\Delta\xi$ as $F_\text{r} = \kappa_\xi \Delta\xi = \kappa_\xi (\xi_\text{r} - \xi_\text{eq})$, where $\kappa_\xi$ is the molecular effective spring constant and $\xi_\text{eq}$ is the electrode gap at which the molecular junction is in mechanical equilibrium ($F=0$). This yields an expression for the probability density function of electrode gaps at rupture given by
\begin{equation}
\label{pr}
    p_\text{r}(\xi_\text{r})=\frac{k_{0\text{r}}\kappa_\xi}{\dot{F}_\text{r}}\exp{\Big[ \beta\chi_\text{r}^\ddag \kappa_\xi(\xi_\text{r} - \xi_\text{eq}) - \frac{k_{0\text{r}}}{\dot{F}_\text{r}\beta\chi_\text{r}^\ddag} \mathrm{e}^{\beta\chi_\text{r}^\ddag \kappa_\xi(\xi_\text{r}-\xi_\text{eq})} \Big]}~.
\end{equation}
Here, all quantities refer to molecular junction rupture and its associated FEP as signaled by the ``$\text{r}$" (junction rupture) subscript throughout. Figure \ref{hist_evol}a shows the $p_0(\xi_0)$ and $p_\text{r}(\xi_\text{r})$ probability density functions for a model system with the set of example parameters shown in Table \ref{tab0}. These parameters were chosen to have values that are representative of break-junction experiments of gold-gold contacts and gold-alkane-gold molecular junctions\cite{huang2006measurement,pobelov2017dynamic}. As  discussed below, the resulting rupture force and conductance histograms obtained from these parameters are within the range of values that are typically measured in experiments.


The probability density function of initial ($p_0(\xi_0)$) and final ($p_\text{r}(\xi_\text{r})$) molecular elongations during pulling determines the probability of visiting a particular molecular junction gap, $\xi$, during the break-junction experiment. To extract this quantity, we assume that for a given trajectory all $\xi$ are equally probable between a given initial and rupture points, in agreement with the constant pulling speed in these experiments and the harmonic picture for the mechanical deformation. The probability of visiting a particular electrode gap during pulling corresponds to the product of the probabilities that the junction has been formed already ($P_0(\xi)$) and has not been ruptured ($1-P_\text{r}(\xi)$) at $\xi$. Therefore, the probability of sampling an electrode gap $\xi$ is
\begin{equation}
\label{Pxi}
\begin{split}
    P(\xi) =& \overbrace{\Big( \int_{-\infty}^{\xi}p_0(\xi_0)d\xi_0 \Big)}^{P_0(\xi)} \overbrace{\Big( 1-\int_{-\infty}^{\xi}p_\text{r}(\xi_\text{r})d\xi_\text{r}  \Big)}^{1-P_\text{r}(\xi)}\\
    =& \Big( 1-\exp \Big[  - \frac{k_{0\text{f}}}{\dot{F}_\text{f}\beta\chi_\text{f}^\ddag} \mathrm{e}^{\beta\chi_\text{f}^\ddag \kappa_\text{f}\xi}   \Big] \Big)\times\\
    &\exp\Big[ - \frac{k_{0\text{r}}}{\dot{F}_\text{r}\beta\chi_\text{r}^\ddag} \mathrm{e}^{\beta\chi_\text{r}^\ddag \kappa_\xi(\xi-\xi_\text{eq})} \Big]~.
\end{split}
\end{equation}
We considered the application of forces in the $(-\infty,\infty)$ range to account for the pushing and pulling of the junction.  Figure \ref{hist_evol}b shows the resulting $P(\xi)$ and its contributions from the probabilities of junction formation and rupture (dashed lines) calculated with Eq. \ref{Pxi} for the parameters in Table \ref{tab0}.  

\subsection{Emergence of conductance histograms}

\begin{figure*}[t!]
 \centering
\includegraphics[scale=1.0]{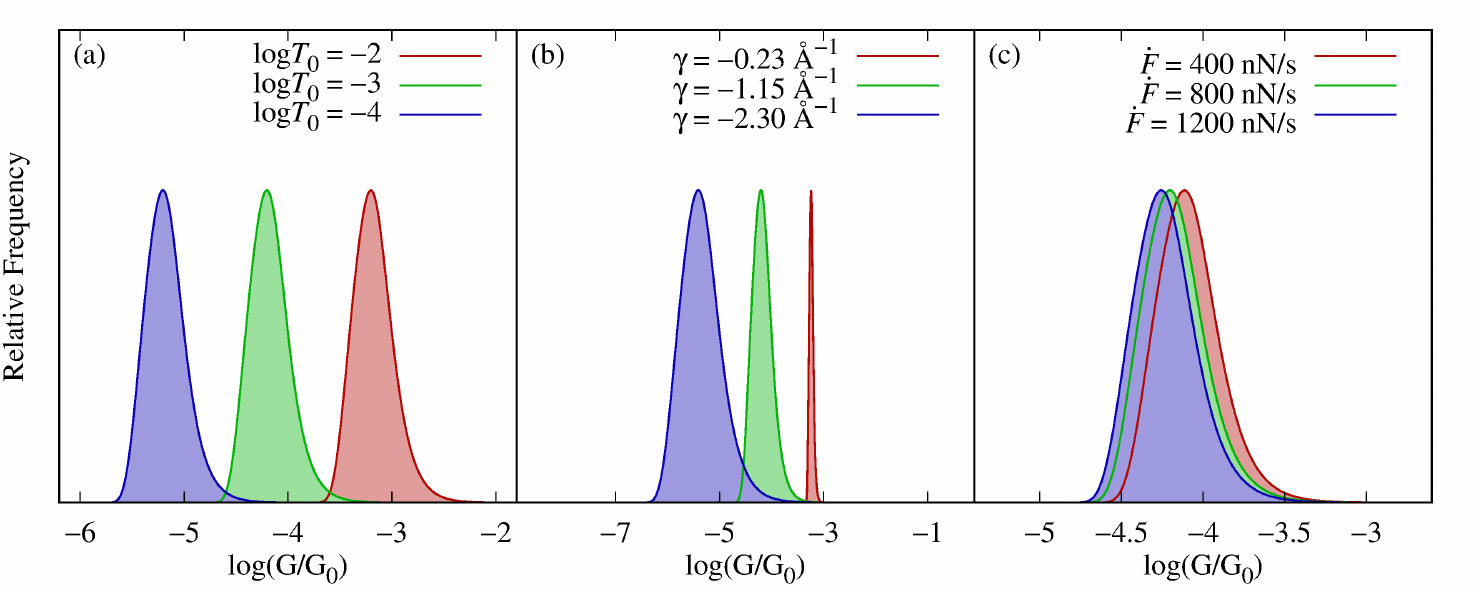}
\caption{(a) Effect of the base transmission, (b) transmission decay coefficient, and (c) loading rate on the break-junction conductance histograms, calculated with Eq. \ref{PlogT}. For these cases, all parameters except the one being varied are those in Table \ref{tab0}. }
\label{hist_param}
\end{figure*}

Equation \ref{Pxi} gives the probability that a given electrode gap is sampled in a break-junction experiment. To recover the conductance histograms, we connect $P(\xi)$ to the probability density of measuring a time-averaged log-transmission $\log T=\log(\mean{G(\xi)}/G_0) \equiv g(\xi)$ (where $G_0=\frac{2e^2}{h}$ is the quantum of conductance) by taking advantage of the (one-to-one) relation between electrode gap and junction conductance. Therefore, Eq. \ref{Pxi} can be rewritten as a probability density for $\log T$ using $\xi = g^{-1}(\log T)$

\begin{equation}
\label{PlogT_gen}
\begin{split}
    p(\log T)
    =& N\Big( 1-\exp \Big[  - \frac{k_{0\text{f}}}{\dot{F}_\text{f}\beta\chi_\text{f}^\ddag} \mathrm{e}^{\beta\chi_\text{f}^\ddag \kappa_\text{f}g^{-1}(\log T)}   \Big] \Big)\times\\
    &\exp\Big[ - \frac{k_{0\text{r}}}{\dot{F}_\text{r}\beta\chi_\text{r}^\ddag} \mathrm{e}^{\beta\chi_\text{r}^\ddag \kappa_\xi(g^{-1}(\log T)-g^{-1}(\log T_\text{eq}))} \Big]~,
\end{split}
\end{equation}
where $\log T_\text{eq}=g(\xi_\text{eq})$ and $N$ is a normalization constant. Equation \ref{PlogT_gen} is an analytical expression for the conductance lineshape in break-junction experiments with a generic dependence between the time-averaged conductance and electrode gap. If the experiments have featureless additive background noise, such as conductance contributions coming from direct electrode-electrode tunneling \cite{gonzalez2006electrical, gotsmann2011direct, quan2015quantitative}, then $p(\log T)_\textrm{EXP} = N_1 p(\log T) + N_0$ where $N_1/N_0$ can be understood as the signal to noise ratio.

Computing a conductance histogram using Eq.~\ref{PlogT_gen} requires a specific form for $g(\xi)$. We now specialize our considerations to the case in which the average transmission $T = T_0\mathrm{e}^{\gamma\xi}$ exponentially increases ($\gamma>0$) or decreases ($\gamma<0$) with the electrode gap. In this case,
\begin{equation}
\label{logT}
    g(\xi)=\log T = \frac{\gamma}{\ln 10}\xi + \log T_0
\end{equation}
is a linear function of $\xi$ with slope determined by $\gamma$. Here, the base transmission $T_0$ is defined as the extrapolated transmission at $\xi\to0$.  This functional dependence was observed in detailed atomistic simulations of alkane-based junctions\cite{li2020understanding}. As discussed below, this specific $g(\xi)$ enables the development of a tractable theory and captures the conductance histograms of a wide class of molecules. Other possible forms can be developed to capture additional features of the conductance histograms.

Substituting Eq. \ref{logT} in Eq. \ref{PlogT_gen} results in an expression for the probability density function of $\log T$
\begin{equation}
\label{PlogT}
    p(\log T) = N\Big( 1-\exp \Big[ -c_2\mathrm{e}^{c_1\log T} \Big] \Big) \exp\Big[ -c_4\mathrm{e}^{c_3\log T} \Big].
\end{equation}
where $c_{1,2}$ and $c_{3,4}$ are characteristic coefficients due to the molecular-junction formation and rupture, respectively, given by 
\begin{equation}
\label{c1}
    c_1 = \frac{\beta\chi_\text{f}^\ddag \kappa_\text{f}}{\gamma}\ln10~,
\end{equation}
\begin{equation}
\label{c2}
    c_2 = \frac{k_{0\text{f}}}{\dot{F}_\text{f}\beta\chi_\text{f}^\ddag} \mathrm{e}^{-c_1 \log T_0}~,
\end{equation}
\begin{equation}
\label{c3}
    c_3 = \frac{\beta\chi_\text{r}^\ddag \kappa_\xi}{\gamma}\ln10,
\end{equation}
and
\begin{equation}
\label{c4}
    c_4 = \frac{k_{0\text{r}}}{\dot{F}_\text{r}\beta\chi_\text{r}^\ddag} \mathrm{e}^{-c_3 \log T_\text{eq}}.
\end{equation}
Here, $\log T_\text{eq} = \frac{\gamma}{\ln 10}\xi_\text{eq} + \log T_0$ is the log-transmission at the equilibrium electrode gap. Figure \ref{hist_evol}c shows the resulting conductance histogram calculated with Eq. \ref{PlogT} for the parameters in Table \ref{tab0}. Equations \ref{PlogT_gen} and \ref{PlogT} are the main results of this section.


\section{Results and discussion}

\begin{figure*}[th!]
 \centering
\includegraphics[scale=1.0]{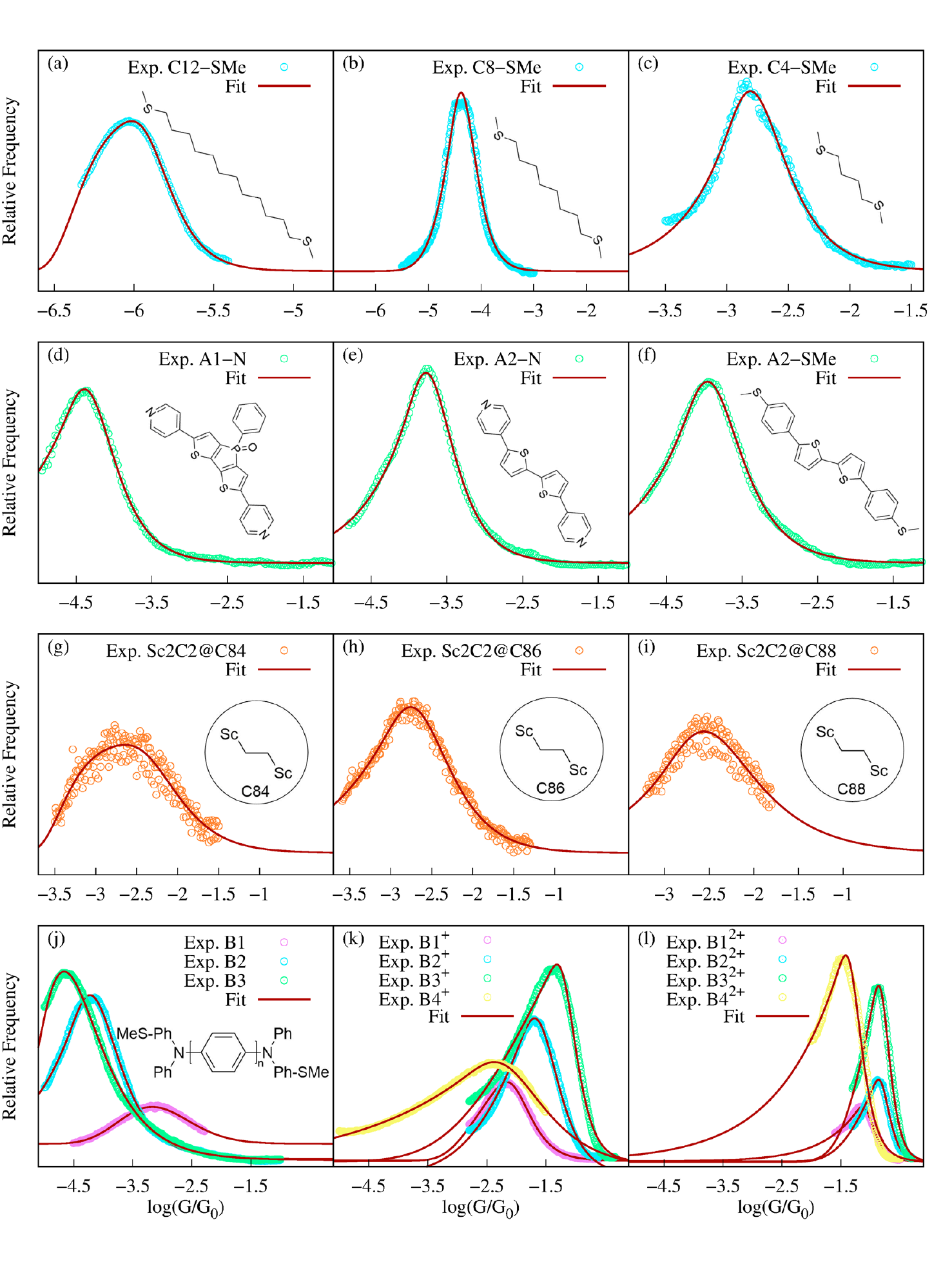}
\caption{Utility of Eq. \ref{PlogT} in fitting experimental conductance histograms. The plot shows experimental conductance histograms of junctions formed with (a)-(c) aliphatic molecules (C$n$-SMe), (d)-(f) aromatic molecules (A$n$-(N or SMe)), (g)-(i) metallofullerene complexes (Sc2C2@C$n$), (j)-(l) radical containing molecules of varying length and charge (B$n^{m+}$) and their accurate fit to Eq. \ref{PlogT}. The values of the fitting parameters are shown in Table \ref{tab1}. Experimental data was provided by Prof. Venkataraman for C$n$-SMe and obtained from Refs.~\cite{daaoub2023not, li2022room, li2022highly} for the other cases.}
\label{hist_exp}
\end{figure*}

Equation \ref{PlogT} provides an analytical expression for the break-junction conductance histograms in terms of physically meaningful parameters. In particular, the coefficients $c_1$ and $c_2$ (Eqs. \ref{c1} and \ref{c2}, respectively) capture the contributions from the metal-metal rupture kinetics that precedes the formation of the molecular junction to the conductance histogram. In turn, parameters $c_3$ and $c_4$ (Eqs. \ref{c3} and \ref{c4}, respectively) capture the contributions from the molecular-junction rupture process. These four coefficients are defined by microscopic parameters describing the free-energy profile of the junction, its mechanical manipulation, and the ability of the molecule to transport charge (see Table~\ref{tab0} for definitions). For completeness, Fig. S1 in the Supporting Information (Appendix A) illustrates the effect of independently varying each coefficient $c_1$-$c_4$ on the conductance histogram.

\subsection{Effect of the microscopic parameters on the conductance histogram}
Equation~\ref{PlogT} enables to elucidate the effects of the microscopic parameters on the conductance histogram. For example, Fig. \ref{hist_param} shows the effect of varying $T_0$, $\gamma$, and $\dot{F}$ (the influence of the remaining parameters is included in Fig. S2). The transmission decay coefficient $\gamma$ and the base transmission $T_0$ are the quantities that define the intrinsic transport properties of the molecule. Figure \ref{hist_param}a shows that $\log T_0$ displaces the distribution of conductances without changing its shape in the logarithmic scale. By contrast, Fig. \ref{hist_param}b shows that the width of the histogram is determined by $\gamma$. Specifically, small values of $|\gamma|$ correspond to molecular junctions  whose conductance is not very sensitive to changes in elongation, resulting in narrow histograms. In the figure, we have covered a wide range of values of $\gamma$ to exemplify this effect; actual experiments might show less dramatic changes. Nevertheless, the chemical design of molecules with small $|\gamma|$ is the key to creating molecular junctions with reproducible conductance features.

Note that the probability distribution of $\xi$ visited once the junction is formed, $P(\xi)$, is asymmetric with a tail toward smaller $\xi$ (see Eq. \ref{Pxi} and  Fig.~\ref{hist_evol}). This asymmetry leads to a tail in the conductance histogram. In our model, negative/positive values of $\gamma$ lead to conductance tails towards the higher/lower conductance values, respectively. Possible additional asymmetries in the conductance histogram that are introduced by background noise in the experiments\cite{zhao2018shaping,gonzalez2006electrical, gotsmann2011direct} need to be removed before inferring the sign of $\gamma$ from experimental data.

The loading rate, $\dot{F} = \kappa\nu$, is proportional to the pulling speed $\nu$ and effective spring constant of the junction $\kappa$ ($\kappa=\kappa_\textrm{f/r}$). Figure \ref{hist_param}c shows that for a molecule with $\gamma<0$, decreasing the loading rate results in the conductance histogram shifting toward higher conductance values. This is because when the junction is elongated slowly, statistically, it breaks at shorter elongations. For $\gamma>0$, decreasing $\dot{F}$ shifts the histograms toward lower conductance values. 

The exponential (or any other monotonic) relation between the electrode gap and conductance  will result in  histograms that depend on the loading rate. This dependency has not been experimentally observed yet in the few characterizations that have been conducted \cite{huang2006measurement, kaliginedi2012correlations, isshiki2020selective}. Under the experimentally realistic conditions of Table~\ref{tab0}, the conductance histogram changes only slightly with loading rate. In fact, varying the loading rate from 400 nN/s to 1200 nN/s only shifts the conductance peak from $\log(\mean{G}/G_0)=-4.11$ to -4.25 (about 0.8 standard deviations of the (log) conductance histogram) which may be challenging to resolve. Therefore, an important challenge for future experiments is to better characterize the dependence of the histograms on loading rate to determine if non-monotonic conductance-electrode gap relations are required to better understand the conductance histograms.

\subsection{Fit to experimental conductance histograms}

To demonstrate that Eq. \ref{PlogT} is useful in analysing experimental data, we tested its ability to fit STM conductance histograms for a wide variety of molecular junctions. Figure \ref{hist_exp} shows the experimental conductance histograms of twenty representative molecular junctions and their excellent fit to Eq. \ref{PlogT}. The extracted parameters are included in Table \ref{tab1}. Even when this is a highly nonlinear fit, we observe that the parameters extracted are robust (see Fig. S3 and Table S1 in the supplementary information, Appendix A). The set includes junctions formed with (a)-(c) aliphatic (SMe)-(CH)$_n$-(SMe)-Au molecules (C$n$SMe), (d)-(f) complex aromatic systems (A1 and A2), (g)-(i) supramolecular complexes composed of metallofullerenes and (j)-(l) radical containing molecules of varying length and charge. Overall, Fig.~\ref{hist_exp} demonstrates the general utility of Eq. \ref{PlogT} to fit experimental histograms.

For completeness, in the supplementary information (Appendix A), we compare the fits using our microscopic theory with the phenomenological approach by Reuter and Ratner in Ref.~\cite{quan2015quantitative,williams2013level} based on Gaussian broadening of energy levels and molecule-metal couplings. While both show good agreement (Fig. S4 and Table S2 in the Supporting Information, Appendix A), the fits to Eq.~\ref{PlogT} are statistically better. Furthermore, our theory has the advantage of being based on a microscopic model that can be used to make physical predictions and advance molecular design.

\begin{figure}[t]
 \centering
\includegraphics[scale=1.0]{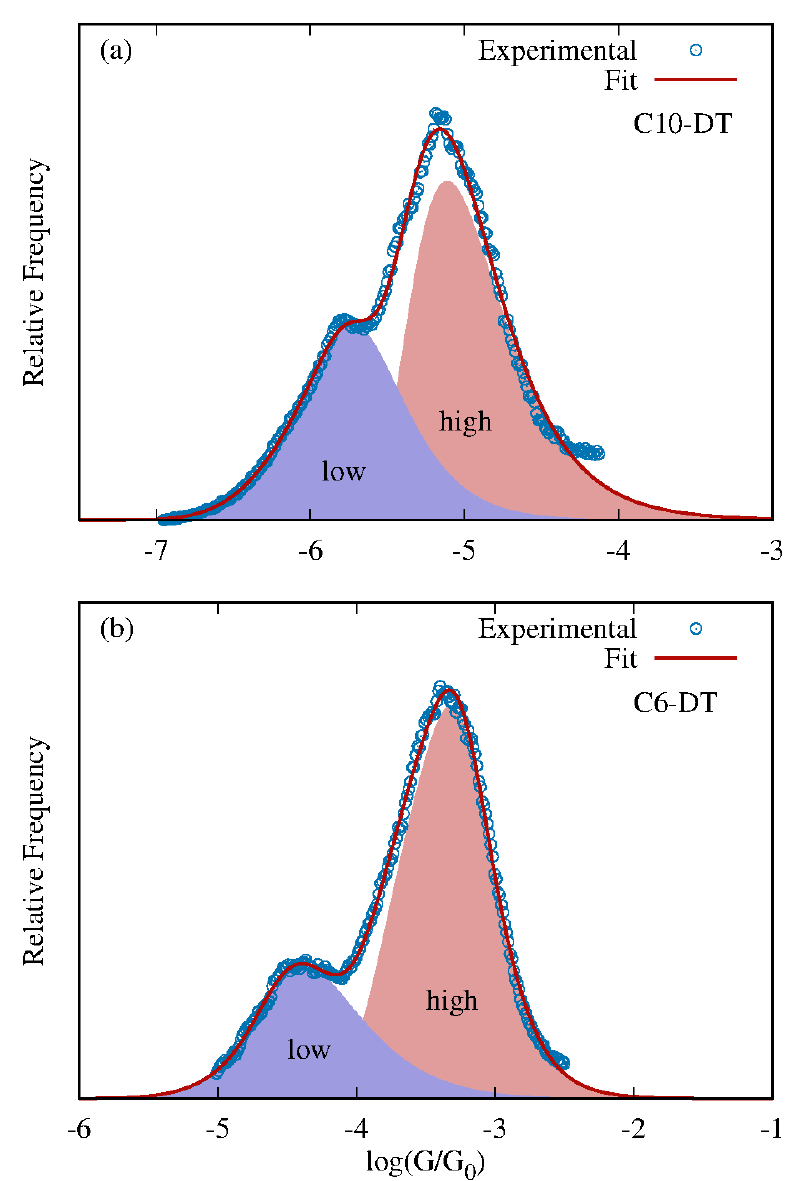}
\caption{Bimodal experimental conductance histograms corresponding to the Au-C$n$-DT-Au junctions and their fit to the $p_1(\log T)+p_2(\log T)$ equation. Here, $p_{1/2}(\log T)$ corresponds to Eq. \ref{PlogT}. The fitting of bimodal distributions allows us to identify individual high/low transmission peaks. The resulting fitting parameters are shown in Table \ref{tab1}. The experimental data was provided by Professor Latha Venkataraman.}
\label{hist_DT}
\end{figure}

Equation \ref{PlogT} can also be used to identify individual contributions to multimodal conductance histograms.  Figure \ref{hist_DT} shows the experimental conductance histogram of two representative Au-S-(CH)$_n$-S-Au junctions (C$n$-DT). This class of molecular junctions leads to bimodal conductance distributions, corresponding to two stable binding configurations of the thiol anchor group on the Au electrode surface (top and bridge configurations)\cite{li2020understanding}. We show that the experimental histogram can be fitted to $p_1(\log T)+p_2(\log T)$. From this fit, individual low and high conductance peaks are then easily identified, as shown in Fig. \ref{hist_DT}. The resulting fitting parameters are included in Table \ref{tab1}.  

Figures \ref{hist_exp} and \ref{hist_DT} further demonstrate that Eq. \ref{PlogT} can be used to capture histograms obtained with both donor-acceptor (-SMe-Au and -N-Au) and covalent (-S-Au) anchoring between the molecule and electrodes.

Figures \ref{hist_exp} and \ref{hist_DT} are based on STM-BJ measurements. Another experimental setup often used to construct conductance histograms is the mechanically-controlled break-junction setup (MCBJ) in a three-point bending configuration\cite{martin2011versatile}. For completeness, we have included fits using Eq.~\ref{PlogT} of MCBJ experiments performed on alkanedithiols and alkanediamines\cite{van2022benchmark} in Fig. S5 and Table S2 of the supplementary information (Appendix A). The excellent fits suggest that the functional form is also applicable in this case. However, additional research on the geometry of forces in MCBJ is required to determine if the force that is parallel to the junction elongation grows linearly in time as required for Bell's model.

Generally, interpreting $c_1-c_4$ directly is challenging since they combine the electrical and mechanical properties of the junction. Below, we describe how to extract the physically interpretable parameters that define the model by supplementing the conductance histograms with force spectroscopy.

\subsection{Extracting microscopic parameters}

To interpret the conductance histograms it is desirable to extract all microscopic parameters that define the $c_1$-$c_4$ coefficients. To do so, it is necessary to complement the conductance measurements with force-spectroscopy of both the metal-metal and the molecular junction. In experiments, the inverse temperature $\beta$, loading rates ($\dot{F}_\text{f}, \dot{F}_\text{r}$) and the elasticity of the junction ($\kappa_\text{f}, \kappa_\xi$) are known.
Fitting the rupture-force histogram of both the metal-metal and molecular junction to Eq. \ref{pF}, yields the spontaneous rupture rate  ($k_0$) and the distance to the transition state ($\chi^\ddag$) in each case. Fitting the conductance histogram to Eq. \ref{PlogT} to extract  $c_1$-$c_4$, and then using the extracted values for the mechanical parameters  in Eqs. \ref{c1}-\ref{c4} yields the conductance decay coefficient ($\gamma$), the molecular base transmission ($\log T_0$), the transmission at mechanical equilibrium ($\log T_\text{eq}$) and its corresponding electrode gap ($\xi_\text{eq}$). This set of parameters completely defines the electro-mechanical model.

To demonstrate this procedure and test its numerical robustness, we generated synthetic rupture force and conductance histograms consistent with the parameters in Table~\ref{tab0} (see SI and  Fig. S3 for details). The extracted microscopic parameters from the synthetic data (column 4, Table ~\ref{tab0}) are in excellent agreement with the original set demonstrating the numerical robustness of the approach.

\begin{table*}[t]
\centering{}\caption{Parameters describing the experimental conductance histograms in Fig. \ref{hist_exp} and \ref{hist_DT} obtained by fitting Eq. \ref{PlogT} and R$^2$ quality of the fit.}\label{tab1}
\begin{tabular}{llllll}
\hline
Molecule & $c_1$ & $c_2$ & $c_3$ & $c_4$ & R$^2$ \\
\hline
C12-SMe &  -7.16 &  $7.82\times10^{-19}$ & -7.04 & $3.07\times10^{-20}$ & 0.998\\
C8-SMe &   -5.72 &  $2.18\times10^{-11}$ & -1.08 & $1.88\times10^{-2}$ & 0.974\\
C4-SMe &   -6.35 &  $2.77\times10^{-8}$ & -0.01 & $2.52\times10^{2}$ & 0.966\\
\hline
A1-N & -4.32  & $1.22\times10^{-8}$  & $-2.86\times10^{-3}$  & $4.39\times10^{2}$  &  0.998 \\
A2-N & -4.84  & $2.15\times10^{-8}$  & $-3.72\times10^{-3}$  & $4.39\times10^{2}$  &  0.996 \\
A2-SMe & -3.95  & $2.82\times10^{-8}$  & $-3.26\times10^{-3}$  & $4.39\times10^{2}$  & 0.992  \\

\hline
Sc2C2@C84 & -2.84 & $2.44\times10^{-3}$ & -4.22 & $4.93\times10^{-7}$ & 0.886\\
Sc2C2@C86 & -3.60 & $9.41 \times10^{-5}$& -0.29& 1.77 & 0.987\\
Sc2C2@C88 & -3.22 & $3.63\times10^{-4}$ & $-7.34\times10^{-2}$& $1.71\times10^{1}$& 0.812\\

\hline
B1 & -5.60 & $5.06\times10^{-12}$  & $-5.02\times10^{-1}$ & 2.27 & 0.994 \\
B2 & -3.55 & $5.48\times10^{-7}$ & $-9.94\times10^{-3}$ & $1.23\times10^{2}$ & 0.998\\
B3 & -1.74 & $6.04\times10^{-7}$ & -2.21 & $2.69\times10^{-5}$ & 0.999 \\

B1$^+$ & -3.94 & $5.25\times10^{-4}$ & -1.81 & $8.51\times10^{-3}$ & 0.999 \\
B2$^+$ & -4.10 & $2.21\times10^{-3}$ & $-4.948\times10^{-1}$ & $9.23\times10^{-1}$ & 0.999 \\
B3$^+$ & -5.36 & $3.04\times10^{-1}$ & $-5.795\times10^{-1}$ & $4.72\times10^{-1}$ & 0.992 \\
B4$^+$ & -2.07 & $1.76\times10^{-2}$ & $-3.28\times10^{-3}$ & $1.49\times10^{2}$ & 0.999 \\

B1$^{2+}$ & -8.64 & $1.50\times10^{-4}$ & $-5.24\times10^{-3}$ & $3.08\times10^{2}$ & 0.994\\
B2$^{2+}$ & -9.01 & $1.16\times10^{-3}$ & -1.19 & $5.03\times10^{-1}$ & 0.998 \\
B3$^{2+}$ & -8.71 & $1.47\times10^{-3}$ & -1.49 & $0.31\times10^{-1}$ & 0.999 \\
B4$^{2+}$ & -6.38 & $5.04\times10^{-4}$ & -2.67 & $3.63\times10^{-3}$ & 0.994 \\

\hline
C10-DT (low) &   -4.53  &  $9.58\times10^{-12}$  &  -1.37  &  $4.48\times10^{-4}$ & 0.995\\
C10-DT (high) &  -3.08  &  $4.31\times10^{-10}$  &  -3.66  &  $6.23\times10^{-9}$ & 0.995\\
C6-DT (low) &  -4.53  &  $1.87\times10^{-9}$  & -0.59 & $3.74\times10^{-1}$ & 0.999\\
C6-DT (high) &  -5.05  & $1.58\times10^{-7}$ &  -3.35  & $3.35\times10^{-6}$ & 0.999\\
\hline
\end{tabular}
\end{table*}


\section{Conclusions}

We developed a rigorous microscopic theory of conductance histograms in molecular electronics by merging the theory of force-spectroscopy  with molecular conductance. As a result, we obtained a general and analytical expression (Eq. \ref{PlogT_gen}) for the break-junction conductance histograms with physically meaningful fitting parameters. Assuming an exponential dependence between the transmission coefficient and electrode gap (Eq. \ref{PlogT}), we obtain practical expressions that provide excellent fits to experimental conductance histograms. The analytical expression has been successfully applied (see Figs.~\ref{hist_exp} and \ref{hist_DT}) to a wide variety of molecules including aliphatic, aromatic, supramolecular, and radical-containing molecules, in junctions with covalent and donor-acceptor anchor groups,  and in cases where the histograms are multi-modal (Figs. \ref{hist_exp}-\ref{hist_DT}).

This theory is based on a physical picture in which the mechanical manipulation of the molecular junction determines the width of the histogram, and the stochastic nature of junction rupture and formation determines its shape. This picture emerged from  a recent theory-experiment analysis of the contributing factors to the conductance histogram that showed that this factor alone could account for the width of conductance events encountered in experiments~\cite{li2020understanding}.

Equation \ref{PlogT} can be used to understand how molecular and mechanical parameters affect the conductance histograms (Fig. \ref{hist_param} and S2). In particular, we showed that the  transmission decay coefficient $\gamma$ determines the conductance width and should be a main parameter to investigate in future works that aim to improve the experimental resolution of conductance measurements. The predictions of Eq.~ \ref{PlogT} can be used to experimentally test the range of validity of the theory. Moreover, Eq. \ref{PlogT} provides clear targets for atomistic modeling that can be used to computationally recover the conductance distributions as needed to establish contact between simulations and experiments.

We further provided a viable experimental strategy to extract all the microscopic parameters that define the mechanical and conductance properties in the proposed model. For this, it is necessary to complement the  conductance histograms with rupture-force histograms for the molecular and metal-metal junctions. Such experiments are needed to test the range of applicability  of the theory and interpret the conductance histogram in terms of the microscopic parameters defined by the proposed model.

The theory is based on a single well in the free energy surface for the molecular junction. This unimodal contribution to the overall conductance histogram can be isolated using machine-learning clustering of experimental data \cite{cabosart2019reference, el2019unravelling, van2022benchmark}. Further, the theory can be  generalized to multi-modal processes in which there is interconversion between different junction configurations that can coexist at a given elongation. This physical situation can be represented through multiple wells in the free energy surface in the presence of force.

Overall, the developments in this paper open the possibility to extract physical information from the conductance histograms characterizing the electro-mechanical microscopic properties of molecular junctions, to design meaningful strategies to tune the conductance histogram, and to help bridge the gap between atomistic simulations and molecular junction experiments.

\section*{Data Availability}
The datasets generated during and/or analysed during the current study are available from the corresponding author on reasonable request.

\section*{Code Availability}
The script used to fit the conductance histogram is included in the Supplementary Information (Appendix A).

\begin{acknowledgments}
We would like to thank Jorge Cossio, Michael Deffner and Carmen Herrmann  for helpful discussions. L.M and I.F. thank the National Science Foundation under Grant No. CHE-1553939 and CHE-2102386 for support. P.C. was supported by MinCiencias, University of Antioquia (Colombia) and the Simons foundation (USA). 
\end{acknowledgments}

\appendix

\renewcommand{\thefigure}{S\arabic{figure}}
\renewcommand{\thetable}{S\arabic{table}}
\setcounter{figure}{0}
\setcounter{table}{0}

\section{Supplementary Information}

\subsection{Effects of parameters $c_1$-$c_4$ on the conductance histogram}

\begin{figure*}[h]
\centering
\includegraphics[scale=0.9]{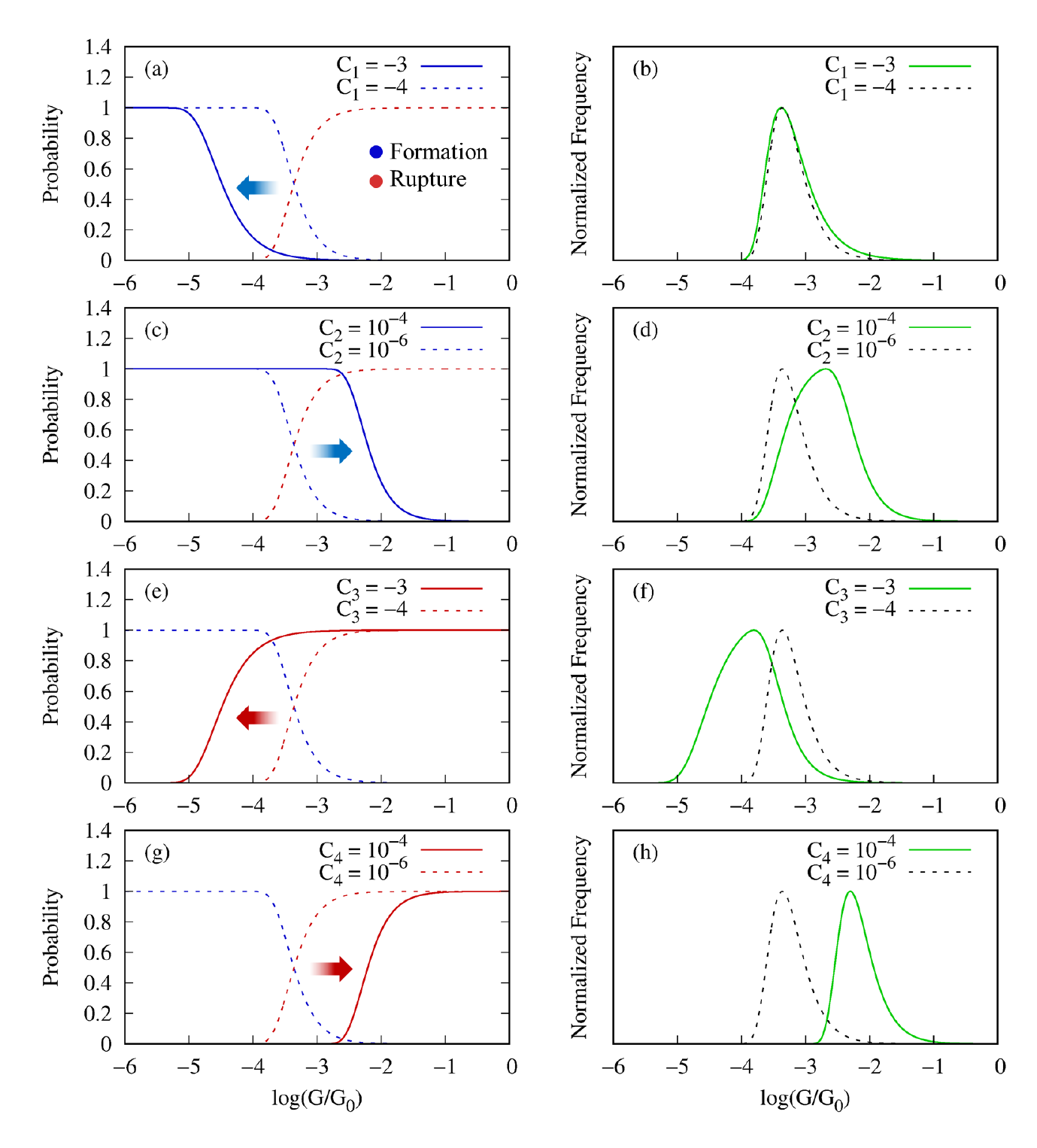}
\caption{Effect of independently varying the fitting parameters $c_1$-$c_4$ on the conductance histogram (Eq. 8). The left panels (a, c, e, g) show the conductance probability considering the junction formation only, i.e. $p(\log T)_\text{f} = ( 1-\exp [ -c_2\mathrm{e}^{c_1\log T}])$ (blue lines), and junction rupture only, i.e. $p(\log T)_\text{r} =\exp[ -c_4\mathrm{e}^{c_3\log T}]$ (red lines). The right panels (b, d, f, h) show the resulting conductance histogram $p(\log T)=p(\log T)_\text{f}\, p(\log T)_\text{r}$, corresponding to Eq. 8. In all cases, we assumed that the conductance decays with the junction gap ($\gamma<0$). Note that, $c_1$ and $c_2$ affect the high-conductance side of the histograms, while parameters $c_3$ and $c_4$ affect the low-conductance side. For $\gamma>0$ the effect is the opposite.}
\label{fig_R4}
\end{figure*}

 To better elucidate the role of $c_1$-$c_4$ in the conductance histograms, in Fig. \ref{fig_R4} we show the effect of independently varying each of the four parameters on the conductance contributions from junction formation and rupture processes (left panels)  and the conductance histograms (right panels). Parameters $c_1$ and $c_2$ control the junction formation (see Fig. \ref{fig_R4}a,c) and therefore affect the higher-conductance side of the histograms (Fig. \ref{fig_R4}b,d) if we assume $\gamma<0$ (i.e. a conductance that decays with the junction gap). By contrast, parameters $c_3$ and $c_4$ control the junction rupture (Fig. \ref{fig_R4}e,g), and therefore, affect the lower-conductance side of the histograms (Fig. \ref{fig_R4}f,h) for $\gamma<0$.

\subsection{Effect of the microscopic parameters on the conductance histogram}

\begin{figure*}[h]
\centering
\includegraphics[scale=1.0]{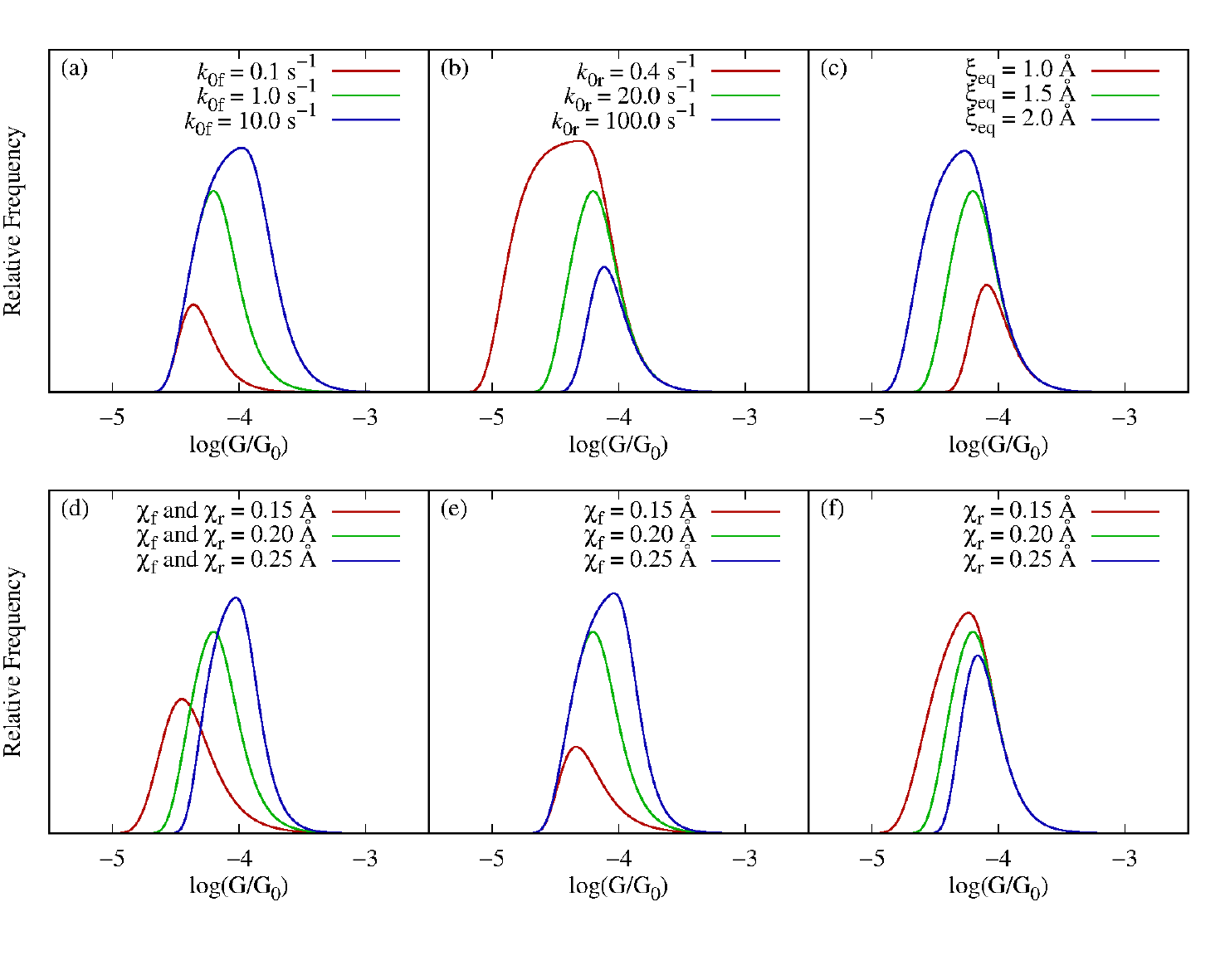}
\caption{Effect of varying the microscopic parameters on the conductance histogram. The panels show the influence of varying (a) metal-metal spontaneous rupture rate; (b) molecular junction spontaneous rupture rate; (c) junction electrode gap at mechanical equilibrium; distance to transition state in the FEP of (d) both the metal-metal contact and molecular junctions, (e) only the metal-metal contact, and (f) only the molecular junction. In all cases, the conductance histogram was calculated using Eq. 8. and the parameters in Table 1 except for the ones being varied.}
\label{fig_par}
\end{figure*}

Figure \ref{fig_par} show the effect of varying the microscopic parameters that define the conductance histogram in our model. This figure complements Fig. 4 in the main text.

Stable metal-metal contacts require the application of higher forces to be ruptured. Thus, metal-metal junctions with smaller spontaneous rupture rates ($k_{0\text{f}}$) lead to longer initial electrode gaps ($\xi_0$). Therefore, as shown in Fig. \ref{fig_par}a, the conductance histogram is shifted towards the low conductance values (for $\gamma<0$) when decreasing $k_{0\text{f}}$. Similarly, a molecular junction with a smaller $k_{0\text{r}}$ will get longer elongated before rupture, allowing the sampling of electrode gaps with an associated lower conductance (for $\gamma<0$) (Fig. \ref{fig_par}b).

The electrode gap at mechanical equilibrium ($\xi_\text{eq}$) indicates how long a junction needs to be elongated before pulling forces are exerted. As shown in Fig. \ref{fig_par}c, a larger $\xi_\text{eq}$ results in conductance histograms with more contributions from low conductance points (for $\gamma<0$). This is because increasing $\xi_\text{eq}$ decreases the force that is being applied to the junction at a given electrode gap, making the junction to rupture at longer elongations.

Finally, changes in the distance from the electrode gap at mechanical equilibrium and the rupture energy barrier ($\chi^\ddag$) effectively change the force-dependent rupture rate (see Eq. 1). Then, varying this parameter causes equivalent trends in the conductance histogram (Fig. \ref{fig_par}d-f) as those observed when varying the spontaneous rupture rate (Fig. \ref{fig_par}a and b).

\subsection{Recovery and robustness of the microscopic parameters from fitting}

Extracting all microscopic parameters that define the coefficients $c_1$-$c_4$ (Eqs. 9-12) requires complementing the conductance measurements with force-spectroscopy of both the metal-metal and the molecular junction. Since this data is not currently available, we demonstrate the procedure and its robustness with synthetic data. 

\begin{figure*}[t]
\centering
\includegraphics[scale=1.0]{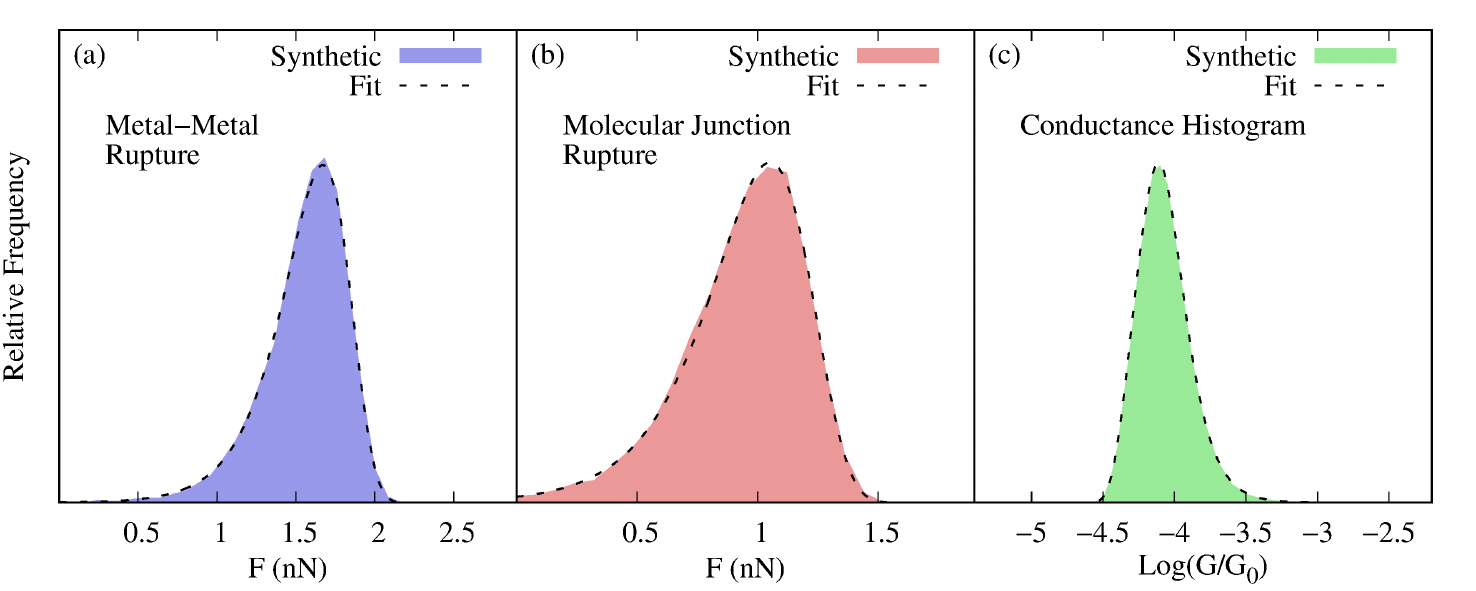}
\caption{Synthetic rupture-force histograms for the (a) metal-metal rupture and (b) molecular junction rupture and their fit to Eq. 2. (c) Numerical conductance histogram and its fit to Eq. 8. The synthetic histograms were generated from 10k pulling trajectories. The parameters resulting from the fitting are shown in tables 1 and \ref{tab:R2}.}
\label{fig_syn}
\end{figure*}

Synthetic data for the rupture force spectroscopy (of both the metal-metal contact and the molecular junction) and for the conductance histogram was generated as follows. We first sampled the probability density functions of initial $p_0(\xi_0)$ (Eq. 3) and rupture $p_r(\xi_r)$ (Eq. 4) electrode gaps to generate corresponding sets of initial $\{\xi_0\}$ and rupture $\{\xi_r\}$ electrode gaps using the parameters in  Table 1. From this data set, the distribution of rupture forces can be reconstructed by taking into account the elastic constant of the metal-metal or molecular junction (taken to be identical to the one of gold  as this is often the softest feature of the junction). The resulting synthetic rupture force histograms are shown in Fig. \ref{fig_syn}a and b. We then selected random pairs from the ($\{\xi_0\}$, $\{\xi_r\}$) set to generate a set of electrode gap elongation trajectories that was then employed to get a set of sampled conductance values (using Eq. 7) and construct the conductance histograms (Fig. \ref{fig_syn}c).

To extract the parameters that summarize the mechanical properties of the junction, we fitted the rupture-force histograms using Eq. 2 for both the metal-metal and molecular junction. From this fit, we extracted the rupture rate at zero force ($k_0$) and the distance to the transition state ($\chi^\ddag$) without using any information about the simulations. In experiments, the inverse temperature $\beta$, loading rates ($\dot{F}_f, \dot{F}_r$) and elasticity of the junctions ($\kappa_f, \kappa_\xi$) are known.

To extract the parameters that summarize the conductance properties of the junction, we employed Eq. 8 to fit the synthetic conductance histograms to extract $c_1$-$c_4$. To test the robustness of the fitting procedure, we compare the extracted parameters to the original parameters in Table \ref{tab:R2} yielding results that are comparable to the original set. Since we now have access to the mechanical parameters, we can now extract the conductance decay coefficient ($\gamma$), the molecular base transmission ($\log T_0$), the transmission at mechanical equilibrium ($\log T_\textrm{eq}$) and its corresponding electrode gap ($\xi_\textrm{eq}$) from $c_1-c_4$. That is, all microscopic parameters can be extracted from two sets of force spectroscopy experiments (one for the metal-metal junction and another one for the molecular junction), and the conductance histogram. Importantly, the extracted parameters exhibit good quantitative agreement with the true original values (see Table 1), showing that the fitting procedure is robust even given that it is highly nonlinear.

In all cases, the fittings were done using the non-linear least squares method, as implemented in the scipy.optimize.curve\_fit python package, as exemplified in the script below:

\begin{widetext}
\noindent import numpy as np\\
from scipy.optimize import curve\_fit\\

\noindent def fit\_func\_F(F,N,f1,f2,N0):\\
    \hspace*{16pt}``Rupture force probability density function, Eq. 2"\\
    \hspace*{16pt}return $\text{N*np.exp(f1*F-f2*np.exp(f1*F)) + N0}$\\

\noindent def fit\_func\_T(g,N,c1,c2,c3,c4,N0):\\
    \hspace*{16pt}``Conductance histogram, Eq. 8"\\
    \hspace*{16pt}return $\text{N*(1-np.exp(-c1*(np.exp(c2*g))))*(np.exp(-c3*(np.exp(c4*g))))+ N0}$\\

\noindent\#Guessed parameters for the rupture force histogram fitting\\
guess\_F = [guess\_N, guess\_f1, guess\_f2, guess\_N0]\\

\noindent\#Rupture force histogram fitting. c\_F contain the fitted parameters and cov\_F the estimated covariance\\
c\_F,cov\_F = curve\_fit(fit\_func\_F,forces\_file,forces\_frequencies\_file,guess\_F)

\noindent\#Guessed parameters for the conductance histogram fitting\\
guess\_T = [guess\_N, guess\_c1, guess\_c2, guess\_c3, guess\_c4, guess\_N0]\\

\noindent\#Rupture force histogram fitting. c\_T contains the fitted parameters and cov\_T the estimated covariance\\
c\_T,cov\_T = curve\_fit(fit\_func\_T,logT\_file,logT\_frequencies\_file,guess\_T)
\end{widetext}

\begin{table}[h]
    \centering
    \begin{tabular}{ccc}
    \hline
      Parameter   &  Original & Fit \\
      \hline
       $c_1$ & -7.74 & -7.76 \\
       $c_2$ & $2.10\times10^{-14}$ & $3.50\times10^{-14}$ \\
       $c_3$ & -7.74 & -7.56 \\
       $c_4$ & $4.20\times10^{-13}$ & $1.29\times10^{-14}$ \\
       \hline
    \end{tabular}
    \caption{Numerical stability of the fit to the conductance histogram. The original values were used to create synthetic conductance histogram that was then fit using Eq. 8. The new extracted parameters are close to the original set and are accurate enough to recover the physical microscopic parameters in Table 1}
    \label{tab:R2}
\end{table}

\subsection{Comparison with the Reuter-Ratner model}

\begin{figure*}[h]
\centering
\includegraphics[scale=1.0]{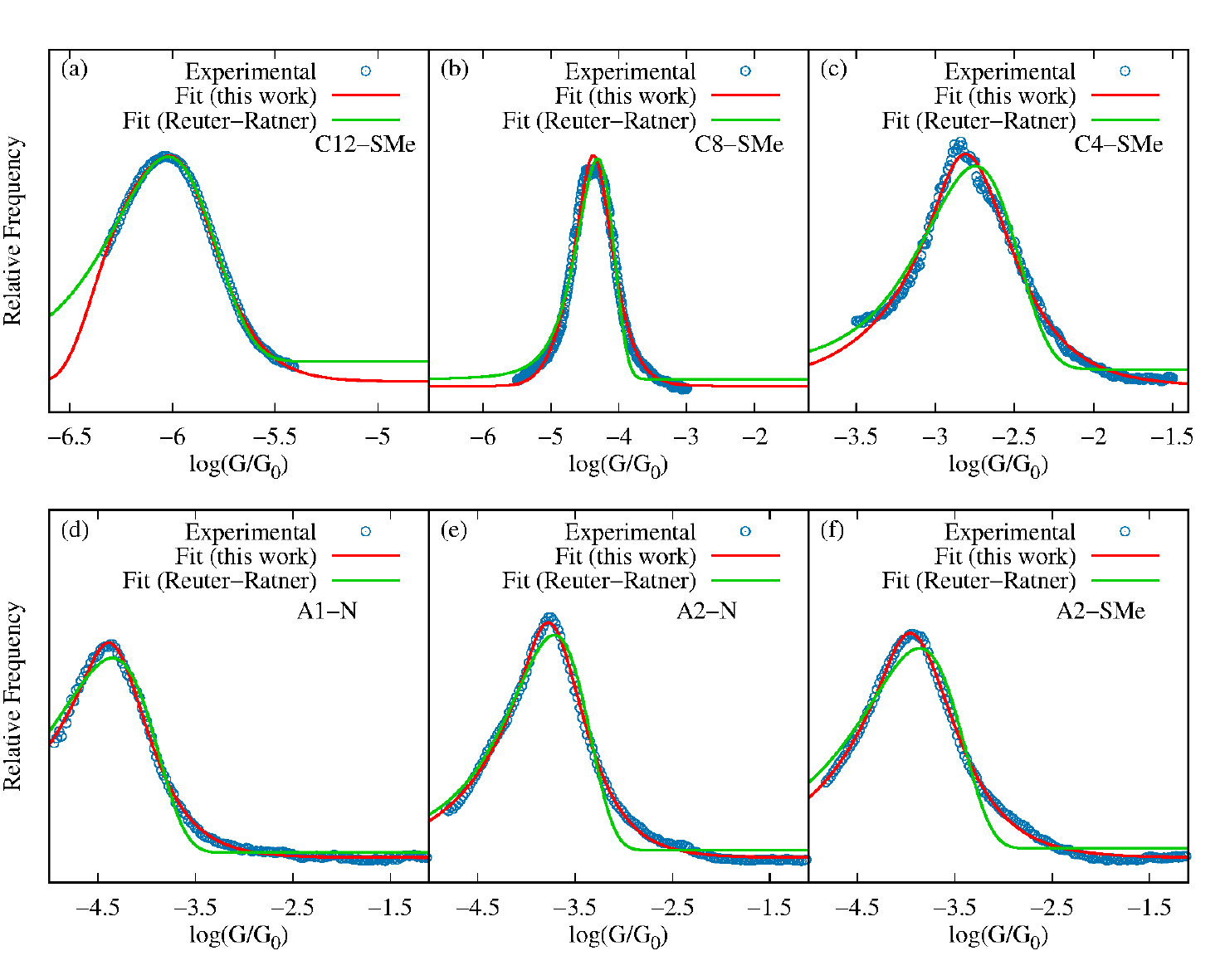}
\caption{Fittings of break-junction conductance histograms in the Cn-SMe series and for the A1-N, A2-N and A2-SMe aromatic molecules using our proposed microscopic theory (Eq. 8) and the phenomenological model by Reuter and Ratner \textit{et al}.\cite{williams2013level,quan2015quantitative}.}
\label{fig_R2}
\end{figure*}

In Fig. \ref{fig_R2}, we contrast fits for the CnSMe series and for the aromatic molecules (A1-N, A2-N and A2-SMe) obtained with the Reuter-Ratner\cite{quan2015quantitative,williams2013level} approach with the ones achievable using the theory in this work. The approach by Reuter and Ratner provides expressions for the conductance histograms based on introducing a phenomenological Gaussian distribution of the level alignment and coupling to the electrodes in the Landauer formula for electron transport. The Reuter-Ratner fits were obtained using Eq. 4 from Ref.~\onlinecite{williams2013level} (equivalent to Eq. 3 in Ref.\onlinecite{quan2015quantitative}) adapted to describe logarithmically binned histograms. The corresponding $\chi^2$ errors and R$^2$ coefficients are shown in Table \ref{tab:R0}. In all cases, both approaches yield reasonable fits of the conductance histograms, with the theory presented in this work yielding  better fits as measured by $\chi^2$ and R$^2$. The definite advantage of our strategy is that the origin of conductance dispersion is linked to microscopic features of the free-energy profile of the junction, its mechanical manipulation, and the ability of the molecule to transport charge. 

\begin{table*}[h]
    \centering
    \begin{tabular}{ccccc}
    \hline
      Molecule   &  R$^2$ (this work) & $\chi^2$ (this work) & R$^2$ (Reuter-Ratner) & $\chi^2$ (Reuter-Ratner) \\
      \hline
        C12-SMe & 0.999  & $6.89\times10^{-5}$   &  0.998  & $1.42\times10^{-4}$  \\
        C8-SMe &  0.993 &  $1.38\times10^{-3}$  &  0.974  & $6.12\times10^{-3}$  \\
        C4-SMe & 0.993  & $1.52\times10^{-3}$ &  0.966  & $8.18\times10^{-3}$  \\
        A1-N & 0.998  & $3.98\times10^{-4}$ & 0.987 & $3.72\times10^{-3}$ \\
        A2-N & 0.996  & $1.55\times10^{-3}$ & 0.974 & $2.58\times10^{-2}$ \\
        A2-SMe & 0.992  & $6.30\times10^{-4}$ & 0.967 & $1.26\times10^{-2}$ \\
       \hline
    \end{tabular}
    \caption{Comparison of the quality of fits for the conductance histograms in the C$n$-SMe series and for the A1-N, A2-N and A2-SMe aromatic molecules using the proposed theory and the Reuter-Ratner approach\cite{williams2013level,quan2015quantitative} as measured by $\chi^2$ errors and R$^2$ coefficients.}
    \label{tab:R0}
\end{table*}

\subsection{Fit to MCBJ experiments}

\begin{figure*}[h]
\centering
\includegraphics[scale=1.0]{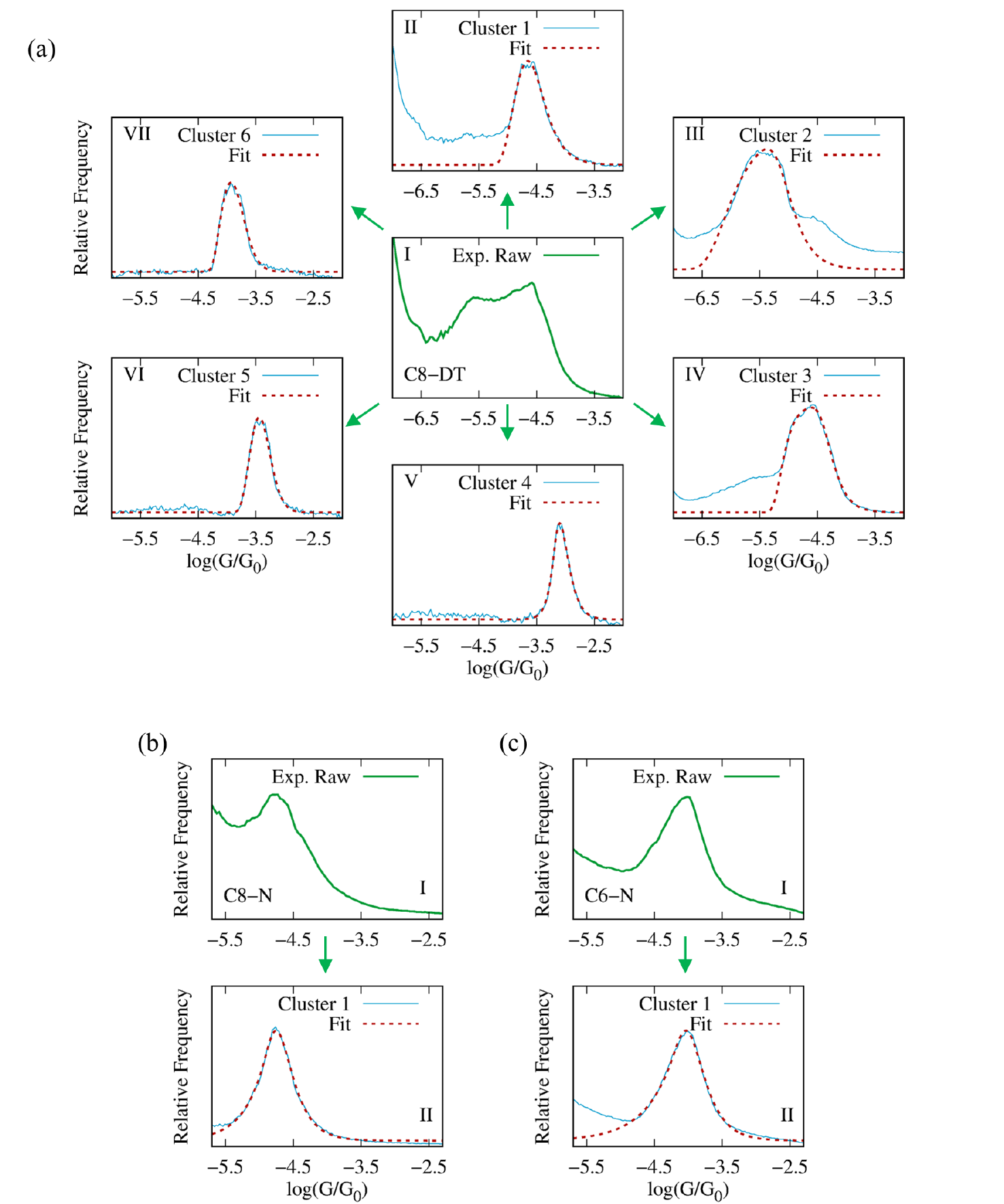}
\caption{MCBJ experimental conductance histograms of junctions containing (a) the C8-DT alkanedithiol and the (b)-(c) C8-N and C6-N alkanediamines and their fit to Eq. 8.  In green is the usual experimental conductance histogram, for which clusters of contributions were isolated using an unsupervised learning algorithm.  In all cases, the experimental data was obtained from Ref. \cite{van2022benchmark}. The excellent fits indicate that the functional form of Eq. 8 is also applicable to MCBJ.}
\label{fig_clus}
\end{figure*}

Figure \ref{fig_clus} shows the fit obtained using  Eq. 8 to MCBJ experiments for C8-DT alkanedithiol and the C8-N and C6-N alkanediamines reported in Ref.\cite{van2022benchmark}. In these experiments, different contributions to the conductance histogram were isolated using an unsupervised learning algorithm, resulting in multiple conductance histograms that can be individually fitted to Eq. 8. The resulting fitting parameters are shown in Table S3.

\begin{table*}[h]
\centering{}\caption{Parameters describing the experimental conductance histograms in Fig. \ref{fig_clus} obtained by fitting to Eq. 8, and R$^2$ quality of the fit.}\label{tab1}
\begin{tabular}{llllll}
\hline
Molecule (cluster) & $c_1$ & $c_2$ & $c_3$ & $c_4$ & R$^2$ \\
\hline
 C8-DT (1)  & -4.64 & $1.00\times10^{-9}$ & -5.98 & $1.507\times10^{-13}$ & 0.996  \\
 C8-DT (2)  &  -4.15 & $8.13\times10^{-10}$ & -3.04 &  $8.52\times10^{-9}$ &  0.970 \\
 C8-DT (3)  &  -5.62 & $3.17\times10^{-11}$ & -7.47 & $2.15\times10^{-17}$ &  0.999  \\
 C8-DT (4)  & $-1.35\times10^{1}$ & $6.94\times10^{-19}$ & -0.85 & $6.86\times10^{-1}$ &  0.999 \\
 C8-DT (5)  &  $-1.70\times10^{1}$ & $1.68\times10^{-28}$ & -1.83 & $1.71\times10^{-2}$ &  0.992 \\
 C8-DT (6)  & $-1.82\times10^{1}$ & $6.28\times10^{-34}$ & -1.25 & $1.09\times10^{-1}$ &  0.986  \\
 C8-N (1)  &  -7.66 & $2.37\times10^{-16}$ &  $-4.54\times10^{-3}$ & $6.94\times10^{2}$ & 0.997  \\
 C6-N (1)  &  -6.66 &  $3.64\times10^{-12}$ & $-6.52\times10^{-3}$ & $3.58\times10^{2}$ &  0.995  \\
\hline
\end{tabular}
\end{table*}

\bibliographystyle{unsrtnat}
\bibliography{references}

\begin{thebibliography}{87}
\providecommand{\natexlab}[1]{#1}
\providecommand{\url}[1]{\texttt{#1}}
\expandafter\ifx\csname urlstyle\endcsname\relax
  \providecommand{\doi}[1]{doi: #1}\else
  \providecommand{\doi}{doi: \begingroup \urlstyle{rm}\Url}\fi

\bibitem[Elke and Carlos(2017)]{elke2017molecular}
Scheer Elke and Cuevas~Juan Carlos.
\newblock \emph{Molecular Electronics: an Introduction to Theory and Experiment}, volume~15.
\newblock World Scientific, 2017.

\bibitem[Datta(2005)]{datta2005quantum}
Supriyo Datta.
\newblock \emph{Quantum transport: atom to transistor}.
\newblock Cambridge university press, 2005.

\bibitem[Nitzan(2006)]{nitzan2006chemical}
Abraham Nitzan.
\newblock \emph{Chemical Dynamics in Condensed Phases: Relaxation, Transfer and Reactions in Condensed Molecular Systems}.
\newblock Oxford university press, 2006.

\bibitem[Coropceanu et~al.(2007)Coropceanu, Cornil, da~Silva~Filho, Olivier, Silbey, and Br{\'e}das]{coropceanu2007charge}
Veaceslav Coropceanu, J{\'e}r{\^o}me Cornil, Demetrio~A da~Silva~Filho, Yoann Olivier, Robert Silbey, and Jean-Luc Br{\'e}das.
\newblock Charge transport in organic semiconductors.
\newblock \emph{Chem. Rev.}, 107\penalty0 (4):\penalty0 926--952, 2007.

\bibitem[Nitzan and Ratner(2003)]{nitzan2003electron}
Abraham Nitzan and Mark~A Ratner.
\newblock Electron transport in molecular wire junctions.
\newblock \emph{Science}, 300\penalty0 (5624):\penalty0 1384--1389, 2003.

\bibitem[Bergfield and Ratner(2013)]{bergfield2013forty}
Justin~P Bergfield and Mark~A Ratner.
\newblock Forty years of molecular electronics: Non-equilibrium heat and charge transport at the nanoscale.
\newblock \emph{Phys. Status Solidi (b)}, 250\penalty0 (11):\penalty0 2249--2266, 2013.

\bibitem[Cabosart et~al.(2019)Cabosart, El~Abbassi, Stefani, Frisenda, Calame, Van~der Zant, and Perrin]{cabosart2019reference}
Damien Cabosart, Maria El~Abbassi, Davide Stefani, Riccardo Frisenda, Michel Calame, Herre~SJ Van~der Zant, and Mickael~L Perrin.
\newblock A reference-free clustering method for the analysis of molecular break-junction measurements.
\newblock \emph{Appl. Phys. Lett.}, 114\penalty0 (14):\penalty0 143102, 2019.

\bibitem[Mej{\'\i}a et~al.(2018)Mej{\'\i}a, Renaud, and Franco]{mejia2018signatures}
Leopoldo Mej{\'\i}a, Nicolas Renaud, and Ignacio Franco.
\newblock Signatures of conformational dynamics and electrode-molecule interactions in the conductance profile during pulling of single-molecule junctions.
\newblock \emph{J. Phys. Chem. Lett.}, 9\penalty0 (4):\penalty0 745--750, 2018.

\bibitem[Venkataraman et~al.(2006{\natexlab{a}})Venkataraman, Klare, Nuckolls, Hybertsen, and Steigerwald]{venkataraman2006dependence}
Latha Venkataraman, Jennifer~E Klare, Colin Nuckolls, Mark~S Hybertsen, and Michael~L Steigerwald.
\newblock Dependence of single-molecule junction conductance on molecular conformation.
\newblock \emph{Nature}, 442\penalty0 (7105):\penalty0 904--907, 2006{\natexlab{a}}.

\bibitem[Mishchenko et~al.(2010)Mishchenko, Vonlanthen, Meded, Burkle, Li, Pobelov, Bagrets, Viljas, Pauly, Evers, et~al.]{mishchenko2010influence}
Artem Mishchenko, David Vonlanthen, Velimir Meded, Marius Burkle, Chen Li, Ilya~V Pobelov, Alexei Bagrets, Janne~K Viljas, Fabian Pauly, Ferdinand Evers, et~al.
\newblock Influence of conformation on conductance of biphenyl-dithiol single-molecule contacts.
\newblock \emph{Nano lett.}, 10\penalty0 (1):\penalty0 156--163, 2010.

\bibitem[Wu et~al.(2020)Wu, Bates, Sangtarash, Ferri, Thomas, Higgins, Robertson, Nichols, Sadeghi, and Vezzoli]{wu2020folding}
Chuanli Wu, Demetris Bates, Sara Sangtarash, Nicolo Ferri, Aidan Thomas, Simon~J Higgins, Craig~M Robertson, Richard~J Nichols, Hatef Sadeghi, and Andrea Vezzoli.
\newblock Folding a single-molecule junction.
\newblock \emph{Nano Lett.}, 20\penalty0 (11):\penalty0 7980--7986, 2020.

\bibitem[Mej{\'\i}a and Franco(2019)]{mejia2019force}
Leopoldo Mej{\'\i}a and Ignacio Franco.
\newblock Force--conductance spectroscopy of a single-molecule reaction.
\newblock \emph{Chem. Sci.}, 10\penalty0 (11):\penalty0 3249--3256, 2019.

\bibitem[Mej{\'\i}a et~al.(2021)Mej{\'\i}a, Garay-Ruiz, and Franco]{mejia2021diels}
Leopoldo Mej{\'\i}a, Diego Garay-Ruiz, and Ignacio Franco.
\newblock Diels--alder reaction in a molecular junction.
\newblock \emph{The Journal of Physical Chemistry C}, 125\penalty0 (27):\penalty0 14599--14606, 2021.
\newblock URL \url{https://doi.org/10.1021/acs.jpcc.1c01901}.

\bibitem[Li et~al.(2017)Li, Haworth, Xiang, Ciampi, Coote, and Tao]{li2017mechanical}
Yueqi Li, Naomi~L Haworth, Limin Xiang, Simone Ciampi, Michelle~L Coote, and Nongjian Tao.
\newblock Mechanical stretching-induced electron-transfer reactions and conductance switching in single molecules.
\newblock \emph{J. Am. Chem. Soc.}, 139\penalty0 (41):\penalty0 14699--14706, 2017.

\bibitem[Aragones et~al.(2016)Aragones, Haworth, Darwish, Ciampi, Bloomfield, Wallace, Diez-Perez, and Coote]{aragones2016electrostatic}
Albert~C Aragones, Naomi~L Haworth, Nadim Darwish, Simone Ciampi, Nathaniel~J Bloomfield, Gordon~G Wallace, Ismael Diez-Perez, and Michelle~L Coote.
\newblock Electrostatic catalysis of a diels--alder reaction.
\newblock \emph{Nature}, 531\penalty0 (7592):\penalty0 88--91, 2016.

\bibitem[Guan et~al.(2018)Guan, Jia, Li, Liu, Wang, Yang, Gu, Su, Houk, Zhang, et~al.]{guan2018direct}
Jianxin Guan, Chuancheng Jia, Yanwei Li, Zitong Liu, Jinying Wang, Zhongyue Yang, Chunhui Gu, Dingkai Su, Kendall~N Houk, Deqing Zhang, et~al.
\newblock Direct single-molecule dynamic detection of chemical reactions.
\newblock \emph{Sci. Adv.}, 4\penalty0 (2):\penalty0 eaar2177, 2018.

\bibitem[Huang et~al.(2017)Huang, Jevric, Borges, Olsen, Hamill, Zheng, Yang, Rudnev, Baghernejad, Broekmann, et~al.]{huang2017single}
Cancan Huang, Martyn Jevric, Anders Borges, Stine~T Olsen, Joseph~M Hamill, Jue-Ting Zheng, Yang Yang, Alexander Rudnev, Masoud Baghernejad, Peter Broekmann, et~al.
\newblock Single-molecule detection of dihydroazulene photo-thermal reaction using break junction technique.
\newblock \emph{Nat. Commun.}, 8:\penalty0 15436, 2017.

\bibitem[Ballmann et~al.(2012)Ballmann, H{\"a}rtle, Coto, Elbing, Mayor, Bryce, Thoss, and Weber]{ballmann2012experimental}
Stefan Ballmann, Rainer H{\"a}rtle, Pedro~B Coto, Mark Elbing, Marcel Mayor, Martin~R Bryce, Michael Thoss, and Heiko~B Weber.
\newblock Experimental evidence for quantum interference and vibrationally induced decoherence in single-molecule junctions.
\newblock \emph{Phys. Rev. Lett.}, 109\penalty0 (5):\penalty0 056801, 2012.

\bibitem[Gu{\'e}don et~al.(2012)Gu{\'e}don, Valkenier, Markussen, Thygesen, Hummelen, and Van Der~Molen]{guedon2012observation}
Constant~M Gu{\'e}don, Hennie Valkenier, Troels Markussen, Kristian~S Thygesen, Jan~C Hummelen, and Sense~Jan Van Der~Molen.
\newblock Observation of quantum interference in molecular charge transport.
\newblock \emph{Nat. Nanotechnol.}, 7\penalty0 (5):\penalty0 305--309, 2012.

\bibitem[Garner et~al.(2018)Garner, Li, Chen, Su, Shangguan, Paley, Liu, Ng, Li, Xiao, et~al.]{garner2018comprehensive}
Marc~H Garner, Haixing Li, Yan Chen, Timothy~A Su, Zhichun Shangguan, Daniel~W Paley, Taifeng Liu, Fay Ng, Hexing Li, Shengxiong Xiao, et~al.
\newblock Comprehensive suppression of single-molecule conductance using destructive $\sigma$-interference.
\newblock \emph{Nature}, 558\penalty0 (7710):\penalty0 415--419, 2018.

\bibitem[Mej{\'\i}a et~al.(2022)Mej{\'\i}a, Kleinekathoefer, and Franco]{mejia2022coherent}
Leopoldo Mej{\'\i}a, Ulrich Kleinekathoefer, and Ignacio Franco.
\newblock Coherent and incoherent contributions to molecular electron transport.
\newblock \emph{The Journal of Chemical Physics}, 156:\penalty0 094302, 2022.

\bibitem[Pirrotta et~al.(2017)Pirrotta, De~Vico, Solomon, and Franco]{pirrotta2017single}
Alessandro Pirrotta, Luca De~Vico, Gemma~C Solomon, and Ignacio Franco.
\newblock Single-molecule force-conductance spectroscopy of hydrogen-bonded complexes.
\newblock \emph{J. Chem. Phys.}, 146\penalty0 (9):\penalty0 092329, 2017.

\bibitem[Koch et~al.(2018)Koch, Li, Nacci, Kumagai, Franco, and Grill]{koch2018structural}
Matthias Koch, Zhi Li, Christophe Nacci, Takashi Kumagai, Ignacio Franco, and Leonhard Grill.
\newblock How structural defects affect the mechanical and electrical properties of single molecular wires.
\newblock \emph{Phys. Rev. lett.}, 121\penalty0 (4):\penalty0 047701, 2018.

\bibitem[Nelson et~al.(2009)Nelson, Kwiatkowski, Kirkpatrick, and Frost]{nelson2009modeling}
Jenny Nelson, Joe~J Kwiatkowski, James Kirkpatrick, and Jarvist~M Frost.
\newblock Modeling charge transport in organic photovoltaic materials.
\newblock \emph{Acc. Chem. Res.}, 42\penalty0 (11):\penalty0 1768--1778, 2009.

\bibitem[Germack et~al.(2009)Germack, Chan, Hamadani, Richter, Fischer, Gundlach, and DeLongchamp]{germack2009substrate}
David~S Germack, Calvin~K Chan, Behrang~H Hamadani, Lee~J Richter, Daniel~A Fischer, David~J Gundlach, and Dean~M DeLongchamp.
\newblock Substrate-dependent interface composition and charge transport in films for organic photovoltaics.
\newblock \emph{Appl. Phys. Lett.}, 94\penalty0 (23):\penalty0 155, 2009.

\bibitem[Breeze et~al.(2001)Breeze, Schlesinger, Carter, and Brock]{breeze2001charge}
AJ~Breeze, Z~Schlesinger, SA~Carter, and PJ~Brock.
\newblock Charge transport in tio 2/m e h- p p v polymer photovoltaics.
\newblock \emph{Phys. Rev. B}, 64\penalty0 (12):\penalty0 125205, 2001.

\bibitem[Yagi and Kaneko(2006)]{yagi2006charge}
Masayuki Yagi and Masao Kaneko.
\newblock Charge transport and catalysis by molecules confined in polymeric materialsand application to future nanodevices for energy conversion.
\newblock \emph{Emissive Materials Nanomaterials}, pages 143--188, 2006.

\bibitem[Heeger(2014)]{heeger201425th}
Alan~J Heeger.
\newblock 25th anniversary article: Bulk heterojunction solar cells: Understanding the mechanism of operation.
\newblock \emph{Adv. Mater.}, 26\penalty0 (1):\penalty0 10--28, 2014.

\bibitem[Golbeck(1992)]{golbeck1992structure}
John~H Golbeck.
\newblock Structure and function of photosystem i.
\newblock \emph{Annu. Rev. Plant Biol.}, 43\penalty0 (1):\penalty0 293--324, 1992.

\bibitem[Sontz et~al.(2012)Sontz, Muren, and Barton]{sontz2012dna}
Pamela~A Sontz, Natalie~B Muren, and Jacqueline~K Barton.
\newblock Dna charge transport for sensing and signaling.
\newblock \emph{Acc. Chem. Res.}, 45\penalty0 (10):\penalty0 1792--1800, 2012.

\bibitem[Franco et~al.(2011)Franco, George, Solomon, Schatz, and Ratner]{Franco-2011}
Ignacio Franco, Christopher~B. George, Gemma~C. Solomon, George~C. Schatz, and Mark~A. Ratner.
\newblock Mechanically activated molecular switch through single-molecule pulling.
\newblock \emph{J. Am. Chem. Soc.}, 133\penalty0 (7):\penalty0 2242--2249, 2011.
\newblock \doi{10.1021/ja1095396}.
\newblock URL \url{https://doi.org/10.1021/ja1095396}.

\bibitem[Li et~al.(2016)Li, Lo, Cai, Zhang, and Yu]{li2016proton}
Lianwei Li, Wai-Yip Lo, Zhengxu Cai, Na~Zhang, and Luping Yu.
\newblock Proton-triggered switch based on a molecular transistor with edge-on gate.
\newblock \emph{Chem. Sci.}, 7\penalty0 (5):\penalty0 3137--3141, 2016.

\bibitem[Ghosh et~al.(2004)Ghosh, Rakshit, and Datta]{ghosh2004gating}
Avik~W Ghosh, Titash Rakshit, and Supriyo Datta.
\newblock Gating of a molecular transistor: Electrostatic and conformational.
\newblock \emph{Nano Lett.}, 4\penalty0 (4):\penalty0 565--568, 2004.

\bibitem[Lang and Solomon(2005)]{lang2005charge}
Norton~D Lang and Paul~M Solomon.
\newblock Charge control in a model biphenyl molecular transistor.
\newblock \emph{Nano letters}, 5\penalty0 (5):\penalty0 921--924, 2005.

\bibitem[Fathizadeh et~al.(2018)Fathizadeh, Behnia, and Ziaei]{fathizadeh2018engineering}
S~Fathizadeh, S~Behnia, and J~Ziaei.
\newblock Engineering dna molecule bridge between metal electrodes for high-performance molecular transistor: An environmental dependent approach.
\newblock \emph{J. Phys. Chem. B}, 122\penalty0 (9):\penalty0 2487--2494, 2018.

\bibitem[D{\'\i}ez-P{\'e}rez et~al.(2009)D{\'\i}ez-P{\'e}rez, Hihath, Lee, Yu, Adamska, Kozhushner, Oleynik, and Tao]{diez2009rectification}
Ismael D{\'\i}ez-P{\'e}rez, Joshua Hihath, Youngu Lee, Luping Yu, Lyudmyla Adamska, Mortko~A Kozhushner, Ivan~I Oleynik, and Nongjian Tao.
\newblock Rectification and stability of a single molecular diode with controlled orientation.
\newblock \emph{Nat. Chem.}, 1\penalty0 (8):\penalty0 635--641, 2009.

\bibitem[Elbing et~al.(2005)Elbing, Ochs, Koentopp, Fischer, von H{\"a}nisch, Weigend, Evers, Weber, and Mayor]{elbing2005single}
Mark Elbing, Rolf Ochs, Max Koentopp, Matthias Fischer, Carsten von H{\"a}nisch, Florian Weigend, Ferdinand Evers, Heiko~B Weber, and Marcel Mayor.
\newblock A single-molecule diode.
\newblock \emph{Proc. Natl. Acad. Sci.}, 102\penalty0 (25):\penalty0 8815--8820, 2005.

\bibitem[Xu and Tao(2003)]{xu2003measurement}
Bingqian Xu and Nongjian~J Tao.
\newblock Measurement of single-molecule resistance by repeated formation of molecular junctions.
\newblock \emph{science}, 301\penalty0 (5637):\penalty0 1221--1223, 2003.

\bibitem[He et~al.(2006)He, Sankey, Lee, Tao, Li, and Lindsay]{he2006measuring}
Jin He, Otto Sankey, Myeong Lee, Nongjian Tao, Xiulan Li, and Stuart Lindsay.
\newblock Measuring single molecule conductance with break junctions.
\newblock \emph{Faraday discuss.}, 131:\penalty0 145--154, 2006.

\bibitem[Widawsky et~al.(2012)Widawsky, Darancet, Neaton, and Venkataraman]{widawsky2012simultaneous}
Jonathan~R Widawsky, Pierre Darancet, Jeffrey~B Neaton, and Latha Venkataraman.
\newblock Simultaneous determination of conductance and thermopower of single molecule junctions.
\newblock \emph{Nano lett.}, 12\penalty0 (1):\penalty0 354--358, 2012.

\bibitem[Konishi et~al.(2013)Konishi, Kiguchi, Takase, Nagasawa, Nabika, Ikeda, Uosaki, Ueno, Misawa, and Murakoshi]{konishi2013single}
Tatsuya Konishi, Manabu Kiguchi, Mai Takase, Fumika Nagasawa, Hideki Nabika, Katsuyoshi Ikeda, Kohei Uosaki, Kosei Ueno, Hiroaki Misawa, and Kei Murakoshi.
\newblock Single molecule dynamics at a mechanically controllable break junction in solution at room temperature.
\newblock \emph{J. Am. Chem. Soc.}, 135\penalty0 (3):\penalty0 1009--1014, 2013.

\bibitem[Venkataraman et~al.(2006{\natexlab{b}})Venkataraman, Klare, Tam, Nuckolls, Hybertsen, and Steigerwald]{venkataraman2006single}
Latha Venkataraman, Jennifer~E Klare, Iris~W Tam, Colin Nuckolls, Mark~S Hybertsen, and Michael~L Steigerwald.
\newblock Single-molecule circuits with well-defined molecular conductance.
\newblock \emph{Nano lett.}, 6\penalty0 (3):\penalty0 458--462, 2006{\natexlab{b}}.

\bibitem[Zhou et~al.(2008)Zhou, Wei, Liu, Chen, Tang, and Mao]{zhou2008extending}
Xiao-Shun Zhou, Yi-Min Wei, Ling Liu, Zhao-Bin Chen, Jing Tang, and Bing-Wei Mao.
\newblock Extending the capability of stm break junction for conductance measurement of atomic-size nanowires: an electrochemical strategy.
\newblock \emph{J. Am. Chem. Soc.}, 130\penalty0 (40):\penalty0 13228--13230, 2008.

\bibitem[Li et~al.(2021)Li, Mej{\'\i}a, Marrs, Jeong, Hihath, and Franco]{li2020understanding}
Zhi Li, Leopoldo Mej{\'\i}a, Jonathan Marrs, Hyunhak Jeong, Joshua Hihath, and Ignacio Franco.
\newblock Understanding the conductance dispersion of single-molecule junctions.
\newblock \emph{J. Phys. Chem. C}, 125\penalty0 (6):\penalty0 3406--3414, 2021.

\bibitem[Gonz{\'a}lez et~al.(2006)Gonz{\'a}lez, Wu, Huber, Van Der~Molen, Sch{\"o}nenberger, and Calame]{gonzalez2006electrical}
M~Teresa Gonz{\'a}lez, Songmei Wu, Roman Huber, Sense~J Van Der~Molen, Christian Sch{\"o}nenberger, and Michel Calame.
\newblock Electrical conductance of molecular junctions by a robust statistical analysis.
\newblock \emph{Nano lett.}, 6\penalty0 (10):\penalty0 2238--2242, 2006.

\bibitem[Li and Franco(2019)]{li2019molecular}
Zhi Li and Ignacio Franco.
\newblock Molecular electronics: Toward the atomistic modeling of conductance histograms.
\newblock \emph{J. Phys. Chem. C}, 123\penalty0 (15):\penalty0 9693--9701, 2019.

\bibitem[Quek et~al.(2009)Quek, Kamenetska, Steigerwald, Choi, Louie, Hybertsen, Neaton, and Venkataraman]{quek2009mechanically}
Su~Ying Quek, Maria Kamenetska, Michael~L Steigerwald, Hyoung~Joon Choi, Steven~G Louie, Mark~S Hybertsen, JB~Neaton, and Latha Venkataraman.
\newblock Mechanically controlled binary conductance switching of a single-molecule junction.
\newblock \emph{Nat. Nanotechnol.}, 4\penalty0 (4):\penalty0 230--234, 2009.

\bibitem[Tschudi and Reuter(2016)]{tschudi2016estimating}
Stephen~E Tschudi and Matthew~G Reuter.
\newblock Estimating the landauer- b{\"u}ttiker transmission function from single molecule break junction experiments.
\newblock \emph{Nanotechnology}, 27\penalty0 (42):\penalty0 425203, 2016.

\bibitem[Paulsson et~al.(2009)Paulsson, Krag, Frederiksen, and Brandbyge]{Paulsson2009a}
Magnus Paulsson, Casper Krag, Thomas Frederiksen, and Mads Brandbyge.
\newblock {Conductance of alkanedithiol single-molecule junctions: A molecular dynamics study}.
\newblock \emph{Nano Lett.}, 9\penalty0 (1):\penalty0 117--121, 2009.
\newblock ISSN 15306984.
\newblock \doi{10.1021/nl802643h}.

\bibitem[Li et~al.(2006)Li, He, Hihath, Xu, Lindsay, and Tao]{Li2006}
Xiulan Li, Jin He, Joshua Hihath, Bingqian Xu, Stuart~M. Lindsay, and Nongjian Tao.
\newblock {Conductance of single alkanedithiols: Conduction mechanism and effect of molecule-electrode contacts}.
\newblock \emph{J. Am. Chem. Soc.}, 128\penalty0 (6):\penalty0 2135--2141, 2006.
\newblock ISSN 00027863.
\newblock \doi{10.1021/ja057316x}.

\bibitem[Li et~al.(2008)Li, Pobelov, Wandlowski, Bagrets, Arnold, and Evers]{Li2008a}
Chen Li, Ilya Pobelov, Thomas Wandlowski, Alexei Bagrets, Andreas Arnold, and Ferdinand Evers.
\newblock {Charge transport in single Au | alkanedithiol | Au junctions: Coordination geometries and conformational degrees of freedom}.
\newblock \emph{J. Am. Chem. Soc.}, 130\penalty0 (1):\penalty0 318--326, 2008.
\newblock ISSN 00027863.
\newblock \doi{10.1021/ja0762386}.

\bibitem[French et~al.(2013)French, Iacovella, Rungger, Souza, Sanvito, and Cummings]{French2013b}
William~R. French, Christopher~R. Iacovella, Ivan Rungger, Amaury~Melo Souza, Stefano Sanvito, and Peter~T. Cummings.
\newblock {Atomistic simulations of highly conductive molecular transport junctions under realistic conditions}.
\newblock \emph{Nanoscale}, 5\penalty0 (9):\penalty0 3654, 2013.
\newblock ISSN 2040-3364.
\newblock \doi{10.1039/c3nr00459g}.
\newblock URL \url{http://xlink.rsc.org/?DOI=c3nr00459g}.

\bibitem[French et~al.(2012)French, Iacovella, and Cummings]{French2012}
William~R. French, Christopher~R. Iacovella, and Peter~T. Cummings.
\newblock {Large-scale atomistic simulations of environmental effects on the formation and properties of molecular junctions}.
\newblock \emph{ACS Nano}, 6\penalty0 (3):\penalty0 2779--2789, 2012.
\newblock ISSN 19360851.
\newblock \doi{10.1021/nn300276m}.
\newblock URL \url{http://pubs.acs.org/doi/abs/10.1021/nn300276m{\%}5Cnhttp://www.ncbi.nlm.nih.gov/pubmed/22335340}.

\bibitem[Quan et~al.(2015)Quan, Pitler, Ratner, and Reuter]{quan2015quantitative}
Robert Quan, Christopher~S Pitler, Mark~A Ratner, and Matthew~G Reuter.
\newblock Quantitative interpretations of break junction conductance histograms in molecular electron transport.
\newblock \emph{ACS nano}, 9\penalty0 (7):\penalty0 7704--7713, 2015.

\bibitem[Reuter et~al.(2012)Reuter, Hersam, Seideman, and Ratner]{reuter2012signatures}
Matthew~G Reuter, Mark~C Hersam, Tamar Seideman, and Mark~A Ratner.
\newblock Signatures of cooperative effects and transport mechanisms in conductance histograms.
\newblock \emph{Nano lett.}, 12\penalty0 (5):\penalty0 2243--2248, 2012.

\bibitem[Williams and Reuter(2013)]{williams2013level}
Patrick~D Williams and Matthew~G Reuter.
\newblock Level alignments and coupling strengths in conductance histograms: the information content of a single channel peak.
\newblock \emph{J. Phys. Chem. C}, 117\penalty0 (11):\penalty0 5937--5942, 2013.

\bibitem[Deffner et~al.(2022)Deffner, Weise, Zhang, Mucke, Proppe, Franco, and Herrmann]{deffner2022learning}
Michael Deffner, Marc~Philipp Weise, Haitao Zhang, Maike Mucke, Jonny Proppe, Ignacio Franco, and Carmen Herrmann.
\newblock Learning conductance: Gaussian process regression for molecular electronics.
\newblock \emph{J. Chem. Theory Comput.}, 2022.

\bibitem[El~Abbassi et~al.(2019)El~Abbassi, Zwick, Rates, Stefani, Prescimone, Mayor, Van Der~Zant, and Duli{\'c}]{el2019unravelling}
Maria El~Abbassi, Patrick Zwick, Alfredo Rates, Davide Stefani, Alessandro Prescimone, Marcel Mayor, Herre~SJ Van Der~Zant, and Diana Duli{\'c}.
\newblock Unravelling the conductance path through single-porphyrin junctions.
\newblock \emph{Chemical science}, 10\penalty0 (36):\penalty0 8299--8305, 2019.

\bibitem[Van~Veen et~al.(2022)Van~Veen, Ornago, Van Der~Zant, and El~Abbassi]{van2022benchmark}
Frederik~H Van~Veen, Luca Ornago, Herre~SJ Van Der~Zant, and Maria El~Abbassi.
\newblock Benchmark study of alkane molecular chains.
\newblock \emph{J. Phys. Chem. C}, 126\penalty0 (20):\penalty0 8801--8806, 2022.

\bibitem[Cossio et~al.(2016)Cossio, Hummer, and Szabo]{cossio2016kinetic}
Pilar Cossio, Gerhard Hummer, and Attila Szabo.
\newblock Kinetic ductility and force-spike resistance of proteins from single-molecule force spectroscopy.
\newblock \emph{Biophys. J.}, 111\penalty0 (4):\penalty0 832--840, 2016.

\bibitem[Hummer and Szabo(2003)]{hummer2003kinetics}
Gerhard Hummer and Attila Szabo.
\newblock Kinetics from nonequilibrium single-molecule pulling experiments.
\newblock \emph{Biophys. J.}, 85\penalty0 (1):\penalty0 5--15, 2003.

\bibitem[Cossio et~al.(2015)Cossio, Hummer, and Szabo]{cossio2015artifacts}
Pilar Cossio, Gerhard Hummer, and Attila Szabo.
\newblock On artifacts in single-molecule force spectroscopy.
\newblock \emph{Proc. Natl. Acad. Sci. U.S.A.}, 112\penalty0 (46):\penalty0 14248--14253, 2015.

\bibitem[Dudko et~al.(2006)Dudko, Hummer, and Szabo]{dudko2006intrinsic}
Olga~K Dudko, Gerhard Hummer, and Attila Szabo.
\newblock Intrinsic rates and activation free energies from single-molecule pulling experiments.
\newblock \emph{Phy. Rev. Lett.}, 96\penalty0 (10):\penalty0 108101, 2006.

\bibitem[Hyeon and Thirumalai(2012)]{hyeon2012multiple}
Changbong Hyeon and D~Thirumalai.
\newblock Multiple barriers in forced rupture of protein complexes.
\newblock \emph{J. Chem. Phys.}, 137\penalty0 (5):\penalty0 055103, 2012.

\bibitem[Evans and Ritchie(1997)]{evans1997dynamic}
Evan Evans and Ken Ritchie.
\newblock Dynamic strength of molecular adhesion bonds.
\newblock \emph{Biophys. J.}, 72\penalty0 (4):\penalty0 1541--1555, 1997.

\bibitem[Bell(1978)]{bell1978models}
George~I Bell.
\newblock Models for the specific adhesion of cells to cells.
\newblock \emph{Science}, 200\penalty0 (4342):\penalty0 618--627, 1978.

\bibitem[Izrailev et~al.(1997)Izrailev, Stepaniants, Balsera, Oono, and Schulten]{izrailev1997molecular}
Sergei Izrailev, Sergey Stepaniants, Manel Balsera, Yoshi Oono, and Klaus Schulten.
\newblock Molecular dynamics study of unbinding of the avidin-biotin complex.
\newblock \emph{Biophys. J.}, 72\penalty0 (4):\penalty0 1568--1581, 1997.

\bibitem[Chen and Springer(2001)]{chen2001selectin}
Shuqi Chen and Timothy~A Springer.
\newblock Selectin receptor--ligand bonds: Formation limited by shear rate and dissociation governed by the bell model.
\newblock \emph{Proc. Natl. Acad. Sci. U.S.A.}, 98\penalty0 (3):\penalty0 950--955, 2001.

\bibitem[Alon et~al.(1995)Alon, Hammer, and Springer]{alon1995lifetime}
Ronen Alon, Daniel~A Hammer, and Timothy~A Springer.
\newblock Lifetime of the p-selectin-carbohydrate bond and its response to tensile force in hydrodynamic flow.
\newblock \emph{Nature}, 374\penalty0 (6522):\penalty0 539--542, 1995.

\bibitem[Tees et~al.(1993)Tees, Coenen, and Goldsmith]{tees1993interaction}
DF~Tees, Olivier Coenen, and Harry~L Goldsmith.
\newblock Interaction forces between red cells agglutinated by antibody. iv. time and force dependence of break-up.
\newblock \emph{Biophys. J.}, 65\penalty0 (3):\penalty0 1318--1334, 1993.

\bibitem[Schmidt et~al.(2012)Schmidt, Filippov, Kersch, Beyer, and Clausen-Schaumann]{schmidt2012single}
Sebastian~W Schmidt, Pavel Filippov, Alfred Kersch, Martin~K Beyer, and Hauke Clausen-Schaumann.
\newblock Single-molecule force-clamp experiments reveal kinetics of mechanically activated silyl ester hydrolysis.
\newblock \emph{ACS nano}, 6\penalty0 (2):\penalty0 1314--1321, 2012.

\bibitem[Lin et~al.(2007)Lin, Chen, Sheng, and Tsao]{lin2007bell}
Han-Jou Lin, Hsuan-Yi Chen, Yu-Jane Sheng, and Heng-Kwong Tsao.
\newblock Bell's expression and the generalized garg form for forced dissociation of a biomolecular complex.
\newblock \emph{Phys. Rev. Lett.}, 98\penalty0 (8):\penalty0 088304, 2007.

\bibitem[Vullev(2005)]{vullev2005modulation}
Velentine~I. Vullev.
\newblock Modulation of dissociation kinetics by external force: Examination of the bell model.
\newblock \emph{Journal of Biological Sciences}, 5\penalty0 (6):\penalty0 744--758, 2005.

\bibitem[Pobelov et~al.(2017)Pobelov, Lauritzen, Yoshida, Jensen, M{\'e}sz{\'a}ros, Jacobsen, Strange, Wandlowski, and Solomon]{pobelov2017dynamic}
Ilya~V Pobelov, Kasper~Primdal Lauritzen, Koji Yoshida, Anders Jensen, G{\'a}bor M{\'e}sz{\'a}ros, Karsten~W Jacobsen, Mikkel Strange, Thomas Wandlowski, and Gemma~C Solomon.
\newblock Dynamic breaking of a single gold bond.
\newblock \emph{Nat. Commun}, 8\penalty0 (1):\penalty0 1--6, 2017.

\bibitem[Huang et~al.(2006)Huang, Xu, Chen, Ventra, and Tao]{huang2006measurement}
Zhifeng Huang, Bingqian Xu, Yuchang Chen, Massimiliano~Di Ventra, and Nongjian Tao.
\newblock Measurement of current-induced local heating in a single molecule junction.
\newblock \emph{Nano lett.}, 6\penalty0 (6):\penalty0 1240--1244, 2006.

\bibitem[Kr{\"u}ger et~al.(2002)Kr{\"u}ger, Fuchs, Rousseau, Marx, and Parrinello]{kruger2002pulling}
Daniel Kr{\"u}ger, Harald Fuchs, Roger Rousseau, Dominik Marx, and Michele Parrinello.
\newblock Pulling monatomic gold wires with single molecules: an ab initio simulation.
\newblock \emph{Phy. Rev. Lett.}, 89\penalty0 (18):\penalty0 186402, 2002.

\bibitem[Wang and Leng(2015)]{wang2015gold}
Huachuan Wang and Yongsheng Leng.
\newblock Gold/benzenedithiolate/gold molecular junction: a driven dynamics simulation on structural evolution and breaking force under pulling.
\newblock \emph{J. Phys. Chem. C}, 119\penalty0 (27):\penalty0 15216--15223, 2015.

\bibitem[Wang et~al.(2021)Wang, Yang, Sun, Liu, Liu, Fu, Wang, and Li]{wang2021decoding}
Shuang-Shuang Wang, Zhi Yang, Feng Sun, Ran Liu, Lin Liu, Xiao-Xiao Fu, Chuan-Kui Wang, and Zong-Liang Li.
\newblock Decoding forming processes of different contact configurations in au-and ag-electrode single-molecule junctions.
\newblock \emph{J. Phys. Chem. C}, 125\penalty0 (49):\penalty0 27290--27297, 2021.

\bibitem[Deffner et~al.(2023)Deffner, Weise, Zhang, Mucke, Proppe, Franco, and Herrmann]{deffner2023learning}
Michael Deffner, Marc~Philipp Weise, Haitao Zhang, Maike Mucke, Jonny Proppe, Ignacio Franco, and Carmen Herrmann.
\newblock Learning conductance: Gaussian process regression for molecular electronics.
\newblock \emph{J. Chem. Theory Comput.}, 19\penalty0 (3):\penalty0 992--1002, 2023.

\bibitem[Gotsmann et~al.(2011)Gotsmann, Riel, and L{\"o}rtscher]{gotsmann2011direct}
B~Gotsmann, H~Riel, and E~L{\"o}rtscher.
\newblock Direct electrode-electrode tunneling in break-junction measurements of molecular conductance.
\newblock \emph{Phys. Rev. B}, 84\penalty0 (20):\penalty0 205408, 2011.

\bibitem[Daaoub et~al.(2023)Daaoub, Morris, B{\'e}land, Demay-Drouhard, Hussein, Higgins, Sadeghi, Nichols, Vezzoli, Baumgartner, et~al.]{daaoub2023not}
Abdalghani Daaoub, James~MF Morris, Vanessa~A B{\'e}land, Paul Demay-Drouhard, Amaar Hussein, Simon~J Higgins, Hatef Sadeghi, Richard~J Nichols, Andrea Vezzoli, Thomas Baumgartner, et~al.
\newblock Not so innocent after all: Interfacial chemistry determines charge-transport efficiency in single-molecule junctions.
\newblock \emph{Angew. Chem. Int. Ed.}, page e202302150, 2023.

\bibitem[Li et~al.(2022{\natexlab{a}})Li, Hou, Yao, Zhang, Wu, Wang, Zhang, Liu, Tang, Wei, et~al.]{li2022room}
Jing Li, Songjun Hou, Yang-Rong Yao, Chengyang Zhang, Qingqing Wu, Hai-Chuan Wang, Hewei Zhang, Xinyuan Liu, Chun Tang, Mengxi Wei, et~al.
\newblock Room-temperature logic-in-memory operations in single-metallofullerene devices.
\newblock \emph{Nat. Mater.}, 21\penalty0 (8):\penalty0 917--923, 2022{\natexlab{a}}.

\bibitem[Li et~al.(2022{\natexlab{b}})Li, Low, Wilhelm, Liao, Gunasekaran, Prindle, Starr, Golze, Nuckolls, Steigerwald, et~al.]{li2022highly}
Liang Li, Jonathan~Z Low, Jan Wilhelm, Guanming Liao, Suman Gunasekaran, Claudia~R Prindle, Rachel~L Starr, Dorothea Golze, Colin Nuckolls, Michael~L Steigerwald, et~al.
\newblock Highly conducting single-molecule topological insulators based on mono-and di-radical cations.
\newblock \emph{Nat. Chem.}, 14\penalty0 (9):\penalty0 1061--1067, 2022{\natexlab{b}}.

\bibitem[Zhao et~al.(2018)Zhao, Liu, Mayer, Coppola, Sun, Kim, Wang, Ni, Chen, Wang, et~al.]{zhao2018shaping}
Zhikai Zhao, Ran Liu, Dirk Mayer, Maristella Coppola, Lu~Sun, Youngsang Kim, Chuankui Wang, Lifa Ni, Xing Chen, Maoning Wang, et~al.
\newblock Shaping the atomic-scale geometries of electrodes to control optical and electrical performance of molecular devices.
\newblock \emph{Small}, 14\penalty0 (15):\penalty0 1703815, 2018.

\bibitem[Kaliginedi et~al.(2012)Kaliginedi, Moreno-Garc{\'i}a, Valkenier, Hong, Garc{\'i}a-Su\'arez, Buiter, Otten, Hummelen, Lambert, and Wandlowski]{kaliginedi2012correlations}
Veerabhadrarao Kaliginedi, Pavel Moreno-Garc{\'i}a, Hennie Valkenier, Wenjing Hong, V{\'i}ctor~M Garc{\'i}a-Su\'arez, Petra Buiter, Jelmer~LH Otten, Jan~C Hummelen, Colin~J Lambert, and Thomas Wandlowski.
\newblock Correlations between molecular structure and single-junction conductance: a case study with oligo (phenylene-ethynylene)-type wires.
\newblock \emph{J. Am. Chem. Soc.}, 134\penalty0 (11):\penalty0 5262--5275, 2012.

\bibitem[Isshiki et~al.(2020)Isshiki, Fujii, Nishino, and Kiguchi]{isshiki2020selective}
Yuji Isshiki, Shintaro Fujii, Tomoaki Nishino, and Manabu Kiguchi.
\newblock Selective formation of molecular junctions with high and low conductance states by tuning the velocity of electrode displacement.
\newblock \emph{Phys. Chem. Chem. Phys.}, 22\penalty0 (8):\penalty0 4544--4548, 2020.

\bibitem[Martin et~al.(2011)Martin, Smit, Egmond, van~der Zant, and van Ruitenbeek]{martin2011versatile}
Christian~A Martin, Roel~HM Smit, Ruud~van Egmond, Herre~SJ van~der Zant, and Jan~M van Ruitenbeek.
\newblock A versatile low-temperature setup for the electrical characterization of single-molecule junctions.
\newblock \emph{Rev. Sci. Instrum}, 82\penalty0 (5), 2011.

\end{thebibliography}

\end{document}